\documentclass[aps,prb,reprint,superscriptaddress]{revtex4-1}
\usepackage{amsmath}
\usepackage{graphicx}
\usepackage{bm}
\usepackage{amssymb}
\usepackage{hyperref}
\usepackage{amsmath}
\usepackage{amssymb}

\begin{document}

\author{Xi-Rong Chen}
\email{These authors contributed equally to this work}
\affiliation{National Laboratory of Solid State Microstructures and department of Physics, Nanjing University, Nanjing, 210093, China}
\affiliation{Collaborative Innovation Center of Advanced Microstructures, Nanjing University, Nanjing 210093, China}

\author{Guangze Chen}
\email{These authors contributed equally to this work}
\affiliation{Department of Applied Physics, Aalto University, 02150, Espoo, Finland}

\author{Yue Zheng}
\affiliation{National Laboratory of Solid State Microstructures and department of Physics, Nanjing University, Nanjing, 210093, China}
\affiliation{Collaborative Innovation Center of Advanced Microstructures, Nanjing University, Nanjing 210093, China}

\author{Wei Chen}
\email{Corresponding author: pchenweis@gmail.com}
\affiliation{National Laboratory of Solid State Microstructures and department of Physics, Nanjing University, Nanjing, 210093, China}
\affiliation{Collaborative Innovation Center of Advanced Microstructures, Nanjing University, Nanjing 210093, China}

\author{D. Y. Xing}
\affiliation{National Laboratory of Solid State Microstructures and department of Physics, Nanjing University, Nanjing, 210093, China}
\affiliation{Collaborative Innovation Center of Advanced Microstructures, Nanjing University, Nanjing 210093, China}

\title{Conductance oscillation in surface junctions of Weyl semimetals}

\begin{abstract}
Fermi arc surface states, the manifestation
of the bulk-edge correspondence
in Weyl semimetals, have attracted much research interest.
In contrast to the conventional Fermi loop, the
disconnected Fermi arcs provide an exotic 2D system
for exploration of novel physical effects
on the surface of Weyl semimetals.
Here, we propose that visible conductance oscillation
can be achieved in the planar junctions fabricated
on the surface of Weyl semimetal with a pair of Fermi arcs.
It is shown that Fabry-P\'{e}rot-type interference inside
the 2D junction can generate conductance oscillation
with its visibility strongly
relying on the shape of the Fermi arcs and their orientation
relative to the
strip electrodes, the latter
clearly revealing the anisotropy
of the Fermi arcs.
Moreover, we show that the visibility
of the oscillating pattern can be
significantly enhanced by a magnetic field
perpendicular to the surface taking advantage of
the bulk-surface connected Weyl orbits. Our work offers
an effective way for the
identification of Fermi arc surface states through transport
measurement and predicts the surface of Weyl semimetal as a novel platform
for the implementation of 2D conductance oscillation.
\end{abstract}
\maketitle

\section{introduction}
Weyl fermion is a massless fermionic particle with definite chirality
named after Hermann Weyl \cite{weyl1929gravitation}, which is
proposed originally as a candidate for fundamental particles.
Though it plays an important role in
quantum theory and the standard model, the verification of Weyl fermion
in high-energy physics remains elusive \cite{RevModPhys.90.015001,RevModPhys.88.030501,RevModPhys.88.030502}.
Recently, Weyl fermion has been observed unexpectedly
in an alternative form as quasi-particle excitations in
a class of condensed matter materials called Weyl semimetals (WSMs) \cite{PhysRevB.83.205101},
thereby inspiring research activities on Weyl physics
and opening a new avenue for exploration of relativistic Weyl fermion
in solid-state physics \cite{Shuichi_Murakami_2007,PhysRevLett.107.127205,PhysRevX.5.011029,
huang2015weyl,PhysRevX.5.031013,
Xu613,xu2015discovery,Xue1501092,
xu2016observation,deng2016experimental,
yang2015weyl,huang2016spectroscopic,PhysRevX.6.031021,
jiang2017signature,belopolski2016discovery,lv2015observation}.
In contrast to its high energy counterpart, the exotic properties of Weyl fermion
in solid-state physics
are usually manifested as anomalous transport and optical phenomena \cite{PhysRevB.86.115133,
PhysRevB.85.241101,PhysRevB.88.104412,PhysRevB.89.081407,
Zhou_2013,Burkov_2015,PhysRevB.92.235205,PhysRevLett.116.077201,PhysRevB.93.085107,
hirschberger2016chiral,PhysRevX.5.031023,shekhar2015extremely,du2016large,
PhysRevB.93.121112,zhang2016signatures}.

One unique feature of the WSMs is the existence of
Fermi arc (FA) surface states at its boundaries \cite{PhysRevB.83.205101},
without any high-energy counterpart. According to the no-go
theorem \cite{NIELSEN198120,NIELSEN1981173}, the Weyl points
in a WSM must appear in pairs with opposite chirality \cite{NIELSEN1983389},
with FA spanning between each pair in the surface Brillouin
zone \cite{PhysRevB.83.205101}. Such disconnected FAs are the
fingerprint of WSMs \cite{huang2015weyl,PhysRevX.5.031013,
Xu613,xu2015discovery,Xue1501092,
xu2016observation,deng2016experimental,yang2015weyl,
huang2016spectroscopic,PhysRevX.6.031021,jiang2017signature,
belopolski2016discovery,lv2015observation}, which play a key role
in the identification
of WSMs in experiment \cite{PhysRevX.5.031013,Xu613,xu2015discovery,
Xue1501092,xu2016observation,deng2016experimental,
yang2015weyl,huang2016spectroscopic,PhysRevX.6.031021,
jiang2017signature,belopolski2016discovery}. Most
experiments on Weyl semimetals
focus on the angle-resolved photoemission spectroscopy,
in which the existence of FAs has been confirmed.
Recent progress have also shown that the
configurations of the FAs
are sensitive to the details of the sample boundary
\cite{Morali1286,yang2019topological,PhysRevB.102.085126},
thus opening the possibility for engineering FAs
and exploring their novel effects and potential applications
through surface device fabrication
and transport measurements. In contrast to
the photoemission spectroscopy experiments,
the surface transport measurement has the advantage of
extracting useful information of the spatially
distribution of the surface states
\cite{PhysRevLett.121.166802,PhysRevB.101.125407,zheng2020andreev}.

\begin{figure}[h]
\centering
\includegraphics[width=0.5\textwidth]{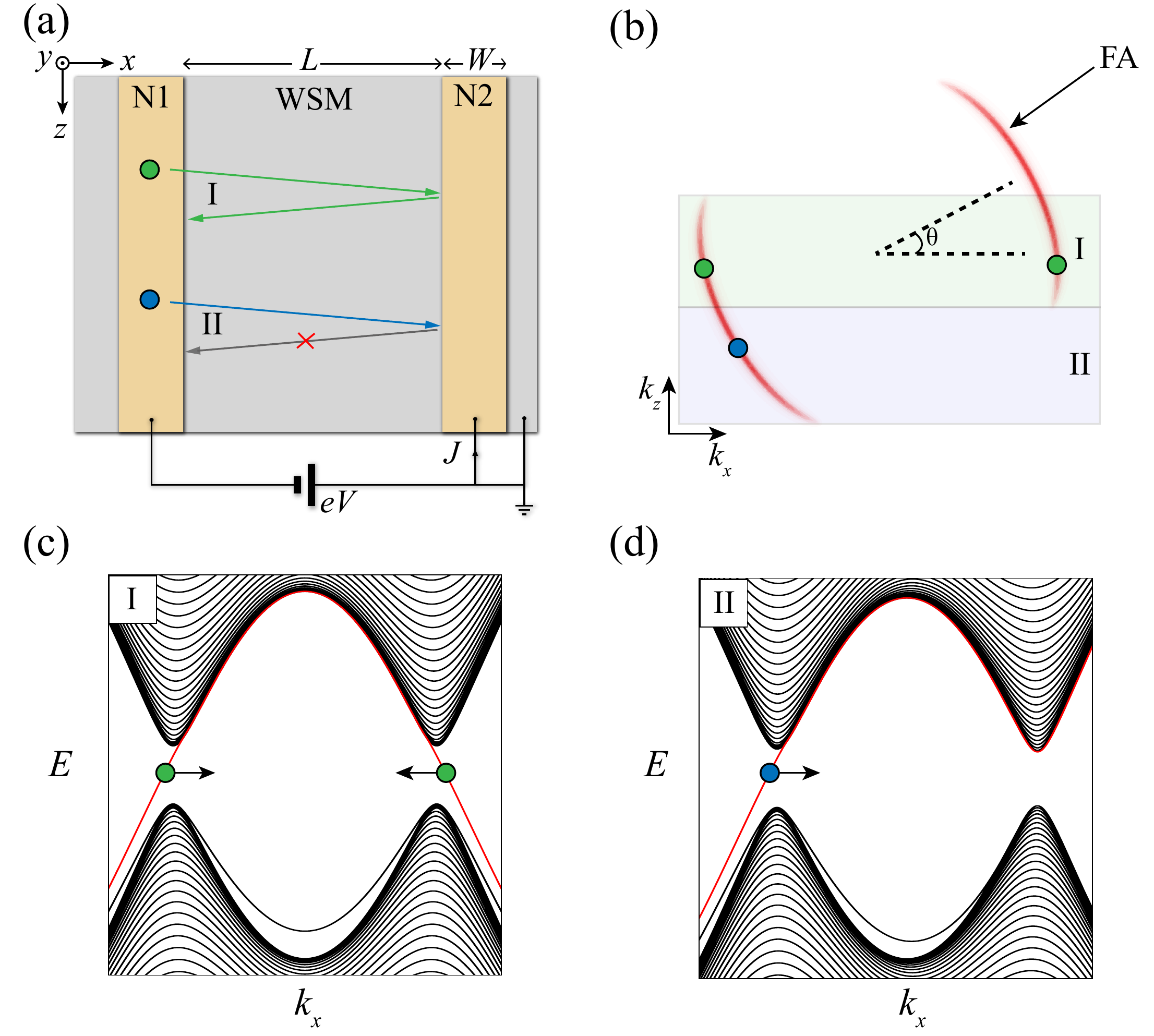}
\caption{(Color online).(a) Schematic illustration of the
planar normal metal (N)-FA-normal metal (N) on the top
surface of a WSM slab and scattering of particles at the
interface. The trajectories of electrons(solid circles)
are sketched as solid lines, where the colors of them
denote that they belong to different $k_z$ regions. (b)
Two regions $\rm{I}$ and $\rm{II}$ for left incident
electrons are defined by the transverse momentum $k_z$.
(c) The corresponding band structures for a fixed $k_z$
of the two regions. The red lines denote the FA top
surface states and black arrows indicate the moving directions.}
\label{fig1}
\end{figure}

In this work, we propose that novel 2D conductance oscillation
stemming from Fabry-P\'{e}rot-type interference
can be realized in the planar normal metal-FA-normal
metal (NFAN) junctions on the surface of the
WSM. The junctions consist of two strip normal
metal electrodes mediated by
a pair of FA surface states in between
as shown in Figs. \ref{fig1}(a) and \ref{fig1}(b).
Our main findings in this work are: (i) Shorter and less curved FAs
can lead to more visible conductance oscillation
stemming from a weaker dephasing effect between different transverse
channels. (ii) The oscillation pattern
of the conductance strongly relies on the relative
orientation between the FAs and the strip electrodes
denoted by the azimuthal angle $\theta$ in Fig. \ref{fig1}(b).
(iii) The visibility of the conductance oscillation
can be significantly enhanced by a magnetic field perpendicular
to the planar junctions due to the existence
of the magnetic Weyl orbit. Our work shows
that FA surface states offer a novel platform
to observe 2D conductance oscillation in addition to
the existing systems such as graphene~\cite{2008Quantum,2013Ballistic,2013A,
Oksanen2014Single,0Quantum,PhysRevLett.113.116601}
and the inverted InAs/GaSb double quantum well~\cite{PhysRevX.10.031007}.
The orientation dependence and the field modulation of the
conductance provide a unique signature of the FAs,
which can be used for identifying
WSMs through transport approach.

The rest of this paper is organized as follows: in Sec. \ref{s2},
we present effective models for a time-reversal ($\mathcal{T}$)-symmetric WSM
and its FA surface states. We then show that a general oriented
FA can be described by applying a rotation transformation of the
effective Hamiltonian. Based on the effective models
and using the Green's function approach, we show
analytically the existence of oscillations in the conductance
spectra of a NFAN
junctions on the WSM surface in Sec. \ref{s3}, and support
the analytical results with numerical simulations
on the lattice model. In Sec. \ref{s4} we show
the dependence of such oscillation on the relative
orientation of the FAs to the normal metals with
numerical means. In Sec. \ref{s5} we show that the
oscillation can be enhanced by applying a magnetic
field perpendicular to the WSM surface. Finally,
we give a brief summary in Sec. \ref{s6}.

\begin{figure*}
\centering
\includegraphics[width=\textwidth]{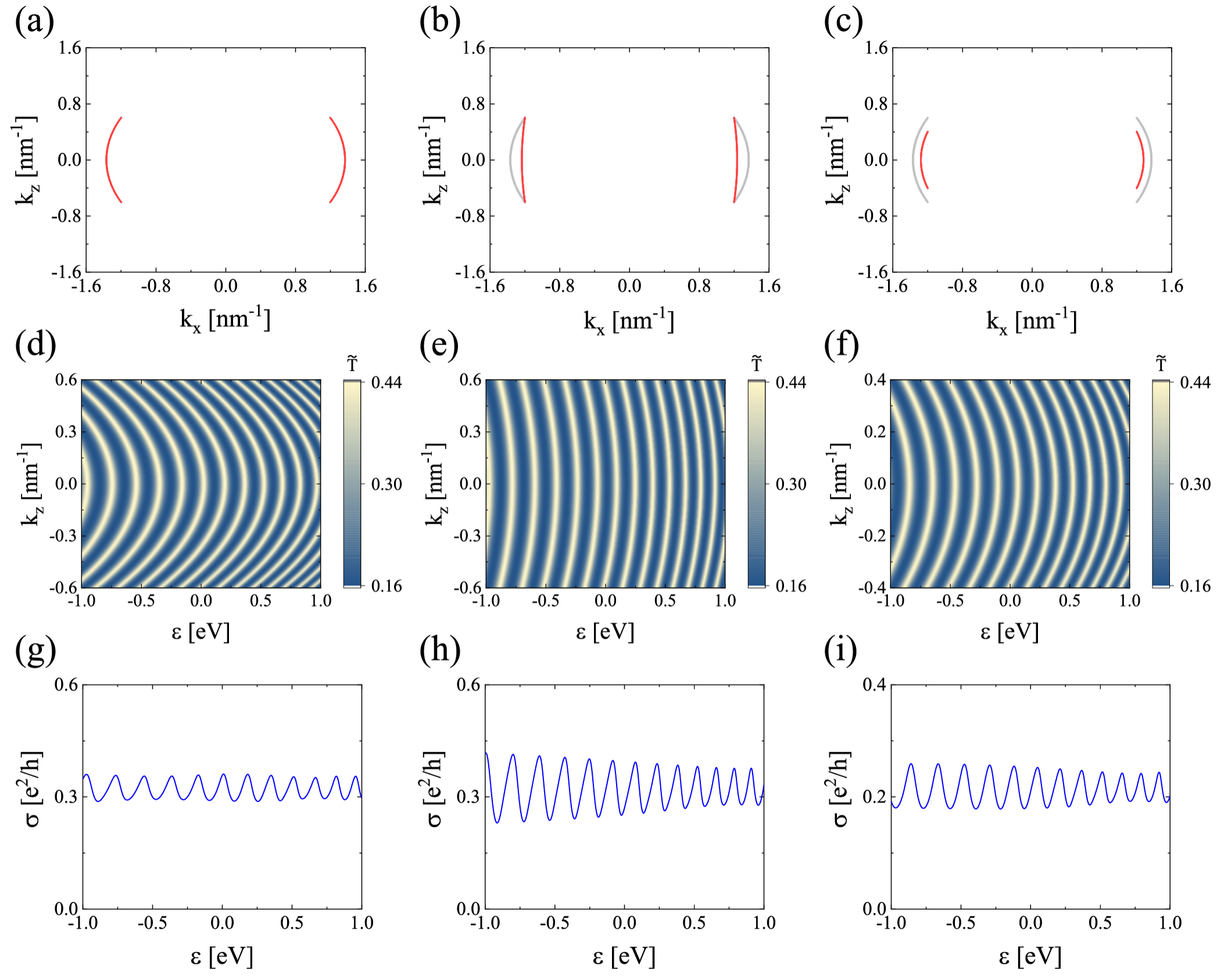}
\caption{(Color online). Top rows illustrate the FAs with different curvature and length, where the relevant parameters are (a) $k_0=0.6\,\rm nm^{-1}$, $d=1.5\,\rm eV\,nm^2$ (b) $k_0=0.6\,\rm nm^{-1}$, $d=0.3\,\rm eV\,nm^2$ and (c) $k_0=0.4\,\rm nm^{-1}$, $d=1.5\,\rm eV\,nm^2$. (d)-(f) Middle rows illustrate the corresponding analytical results of the transmission coefficient $\widetilde{T}(\theta=0)$ as a function of $\varepsilon$ and $k_z$. (g)-(i) Bottom rows exhibit the corresponding conductance $\sigma$ after integrates over $k_z$. The other parameters are set as $a=1 \rm nm$, $R_1=R_2=1$, $M_1=1.25\,\rm eV\,nm^2$ and $k_1=1.2\,\rm nm^{-1}$.}
\label{fig2}
\end{figure*}

\section{$\mathcal{T}$-symmetric Weyl semimetal and FA surface states}\label{s2}
We adopt the following effective two-band $k\cdot p$ model, which describes a $\mathcal{T}$-symmetric Weyl semimetal with four Weyl points \cite{PhysRevB.101.125407,zheng2020andreev}:
\begin{equation}\label{H1}
H_W^0(\bm k)=M_1(k^2_1-k^2_x)\sigma_x+v_yk_y\sigma_y+M_2(k^2_0-k^2_y-k^2_z)\sigma_z,
\end{equation}
where $v_y$ is the velocity in the $\hat y$ direction, $M_{1,2}$
and $k_{0,1}$ are model parameters, $\sigma_{x,y,z}$ are the Pauli matrices
in the pseudo-spin space. The valence and conduction bands cross
linearly at four Weyl points $\bm{k}_W=(\pm k_1,0,\pm k_0)$.
The low-energy Hamiltonian near the Weyl points are
$
h_W(\bm k)=\pm 2M_1k_1k_x\sigma_x+v_yk_y\sigma_y\pm 2M_2k_0k_z\sigma_z.
$
We are interested in the topologically-protected FA surface states on
the open surface in the $\hat y$ direction.
They are confined by $|k_z|<k_0$ and can be described by
\begin{equation}
H_{\text{Arc}}^0(k_x)=M_1(k_x^2-k_1^2),
\end{equation}
with two straight FAs defined by $k_x=\pm k_1$. Generically, FAs in real materials are curved,
we introduce a dispersion term $\epsilon(k_z)=d(k_z^2-k_0^2)$ to capture that feature
and the total Hamiltonian of the surface states are
\begin{equation}\label{FA}
H_{\text{Arc}}(\bm k)=H_{\text{Arc}}^0(k_x)+\epsilon(k_z),
\end{equation}
with the in-plane wave vector $\bm k=(k_x, k_z)$.

One important feature of the FAs is its strong anisotropy.
In the planar junctions, the relative orientation between
the FAs and the normal of the strip electrodes denoted by the angle
$\theta$ [Fig. \ref{fig1}(b)] strongly affects the physical results.
In the long wave-length limit,
we can apply a rotational transformation to the effective Hamiltonian
Eq. \eqref{H1} of WSM to describe such an effect
while fixing the direction
of the electrodes at the same time.
A rotation about the $\hat{y}$-axis by an angle $\theta$ is described by
 \begin{equation} \label{eq5}
     H_W(\bm k,\theta)=H_W^0(U_y^{-1}\bm k)
 \end{equation}
 with the rotation operator \cite{Chen_2013}

\begin{equation}
U_y(\theta)=\left({\begin{matrix}
	\cos\theta& 0 & \sin\theta\\
	0          & 1 & 0          \\
	-\sin\theta& 0 & \cos\theta
\end{matrix}}\right).
\end{equation}
The locations of Weyl points determined by $H_W(\bm k,\theta)=0$ are transferred to
$U_y(\theta)\bm{k}_W=\pm(k_1\cos\theta+k_0\sin\theta,0,-k_1\sin\theta+k_0\cos\theta)$
and the FAs terminated at these points rotate accordingly [cf. Fig. \ref{fig1}(b)].
In the next section, we show the conductance oscillation in
the planar junctions with the dispersion
\eqref{FA}, and in Sec. \ref{s4} we show the dependence
of such oscillation on the orientation of the FAs based on
the discrete version of Hamiltonian \eqref{eq5}.

\section{Conductance oscillation in NFAN junctions}\label{s3}

The WSM surface with FA states
is a novel 2D system, which differs from
other systems with closed Fermi loops.
The disconnected nature of FA may lead to the absence of back-scattering channels
in surface transport \cite{zheng2020andreev}. In particular, consider the NFAN
junctions as shown in Fig. \ref{fig1}(a) with the FA having an azimuthal angle $\theta$
relative to the normal
metal electrodes [Fig.\ref{fig1}(b)]. In region I of the surface Brillouin
zone, there exists two counter-propagating channels at the Fermi
surface [Fig. \ref{fig1}(c)], thus enabling back-scattering;
In contrast, in region II, there exists a single chiral
channel [Fig. \ref{fig1}(d)], and back-scattering is prohibited.
Interestingly, the ratio between region I and region II
depends solely on the relative orientation $\theta$. We
show first the existence of conductance oscillation
with $\theta=0$ in this section, and investigate
its $\theta$ dependence in the next section.

\subsection{Analytical calculation}
We investigate the ballistic transport
in the NFAN junctions using the Green's function method. We
consider the case $\theta=0$ first, where all conducting channels are of type
I in the surface Brillouin zone [cf. Fig. \ref{fig1}]. The surface
Hamiltonian is captured by $H_{\text{Arc}}(\bm k)$
in Eq. \eqref{FA}. The tunneling Hamiltonian
\begin{equation}\label{T}
H_T=\sum_{p,\alpha=1,2}T_{\alpha}d_{p,\alpha}^{\dagger}\psi(x_{\alpha})+H.c.,
\end{equation}
is adopted to describe the coupling between the FA
surface states and the normal electrods
where $T_{\alpha}$ is the tunneling strength between the
surface states and the $\alpha$-electrode [cf. Fig. \ref{fig1}(a)], $d_{p,\alpha}$ is
the Fermi operator in the $\alpha$-terminal with
momentum $p$, and $\psi(x_{\alpha})$ is the field
operator of the surface states at each terminal located at $x_{\alpha}$.

For the planar junctions with good quality of
the strip electrodes the transverse momentum $k_z$
is approximately conserved during scattering.
The differential conductance (without spin degeneracy)
per unit length of the strip electrodes is
the summation over transmissions in all $k_z$ channels as
\begin{equation}\label{cond}
    \sigma(\varepsilon)=\frac{e^2}{h}\int_{-k_0}^{k_0}\widetilde{T}(\varepsilon,k_z)dk_z,
\end{equation}
with $\varepsilon$ the electron energy.
The range of integration is limited by the
spreading of FAs in the $k_z$ direction.
The $k_z$-dependent transmission function
$\widetilde{T}(\varepsilon,k_z)$ can be calculated by the non-equilibrium Green's
function method through \cite{datta1997electronic}
\begin{equation}\label{TT}
\begin{aligned}
\widetilde{T}(\varepsilon,k_z)={\rm Tr}[\Gamma_1 G^R \Gamma_2 G^A],
\end{aligned}
\end{equation}
where $\Gamma_{1,2}$ are linewidth functions of the leads
and $G^{R,A}$ are the full retarded/advanced Green's function.
For a given energy $\varepsilon$ and transverse momentum
$k_z$, there are two counter-propagating channels
with momenta $\pm k_x^0$ and $k_x^0(\varepsilon, k_z)=\sqrt{k_1^2+\frac{\varepsilon-d(k_z^2-k_0^2)}{M_1}}$.
The bare Green's functions can be obtained as
\begin{equation}\label{gr1}
\begin{aligned}
{g}_{\varepsilon}^R(x',x)&=[{g}_{\varepsilon,k_z}^A(x,x')]^*={g}_{\varepsilon,k_z}^R(x,x')\\
&=-\pi i\rho_S(\varepsilon,k_z) e^{ik_0(x'-x)},
\end{aligned}
\end{equation}
with $\rho_S=\frac{1}{4\pi M_1k_x^0}$.
The full Green's function and the linewidth function
can be calculated in the standard way by
taking into account the tunneling term $H_T$,
which gives
\begin{equation}
\begin{split}
{G}_{\varepsilon}^{R}(x_2,x_1)&=\frac{{g}_{\varepsilon}^{R}(x_2,x_1)}{(1+R_1)(1+R_2)-R_1R_2f_{\varepsilon}(x_2,x_1)},\\
\Gamma_{\alpha}(x, x', \varepsilon)&=2\pi\rho_{\alpha}(\varepsilon)|T_{\alpha}|^2\delta(x-x_{\alpha})\delta(x'-x_{\alpha}).
\end{split}
\end{equation}
where $f_{\varepsilon}(x_2,x_1)= e^{2ik_0(x_2-x_1)}$,
$R_{\alpha}(\varepsilon)=\pi^2\rho_S(\varepsilon)\rho_{\alpha}(\varepsilon)|T_{\alpha}|^2$ and $\rho_\alpha$
is the density of states of the leads. We have assumed that
$T_\alpha$ and $\rho_\alpha$ are $k_z$-independent such that
$\Gamma_\alpha$ has no $k_z$ dependence. We have also neglected the $k_z$ dependence of $\rho_S$ that does not qualitatively change the result. The transmission
coefficient in Eq. \eqref{TT} reduces to

\begin{align}
\widetilde{T}(\varepsilon,k_z)&=\frac{4R_1R_2}{|(1+R_1)(1+R_2)-R_1R_2f_{\varepsilon}(x_2,x_1)|^2}.
\end{align}
The Fabry-P\'{e}rot-type interference is indicated
by the coherence factor $f_{\varepsilon}(x_2,x_1)$ in the
transmission function, which induces
the oscillation of $\widetilde{T}$
with varying $\varepsilon$.
It also exhibits a $k_z$ dependence, meaning
that different transverse channels can have a
relative phase shift; see Fig. \ref{fig2}. From the expression
of the conductance \eqref{cond}, one can
infer that a strong dephasing between
the $k_z$ channels will suppress
the overall oscillation of the conductance
by phase averaging. This is the main reason
why the Fabry-P\'{e}rot oscillation of the conductance
in a 2D metal is hard to implement \cite{PhysRevX.10.031007}.
In contrast to a closed Fermi surface, the terminated FAs can
effectively reduce the dephasing effect, so that
the FA surface states provide a promising 2D
platform to implement conductance oscillation.
From the physical picture above,
we can infer that FAs with smaller curvature and
shorter length result in more visible oscillation,
that is verified in Fig. \ref{fig2}.

\subsection{Numerical simulation}\label{s3b}

\begin{figure*}
\centering
\includegraphics[width=\textwidth]{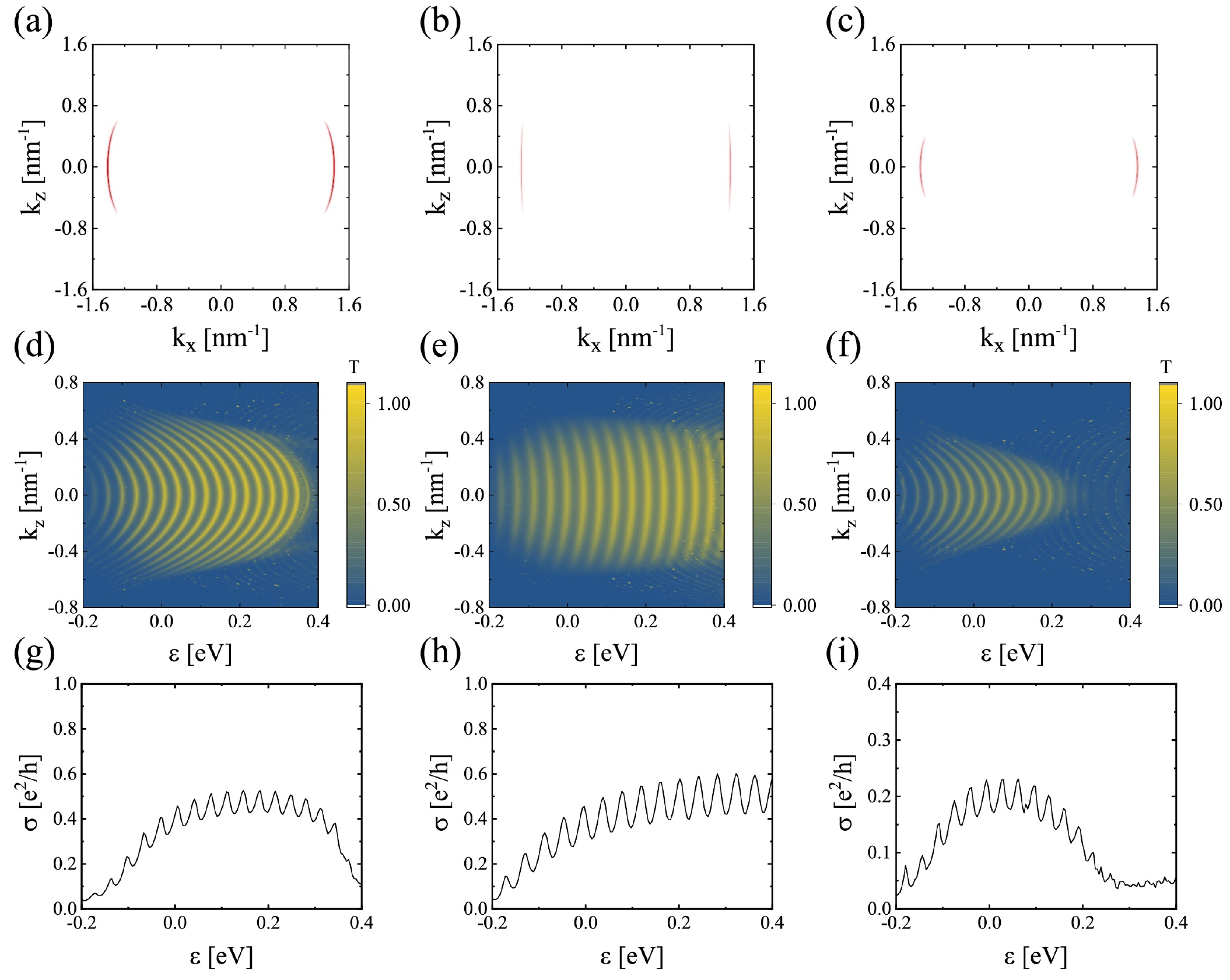}
\caption{(Color online). Top rows illustrate the FA spectra with different curvature and length, where the relevant parameters are (a) $k_0=0.6\,\rm nm^{-1}$, $V=1.0\,\rm eV$ (b) $k_0=0.6\,\rm nm^{-1}$, $V=0.2\,\rm eV$ and (c) $k_0=0.4\,\rm nm^{-1}$, $V=1.0\,\rm eV$. (d)-(f) Middle rows illustrate the corresponding numerical results of the transmission coefficient $T$ as a function of $\varepsilon$ and $k_z$. (g)-(i) Bottom rows exhibit the corresponding conductance $\sigma$ after integrates over $k_z$. The other parameters are $a=1\, \rm nm$, $U_1=U_2=1.0\,\rm eV$, $M_1=M_2=1.25\, \rm eV\,nm^2$, $v_y=0.66\,\rm eV\,nm$, $k_1=1.2\,\rm nm^{-1}$, $C=0.5\,\rm eV\,nm^2$, $\mu_N=1.0\,\rm eV$, and $t_N=0.5\,\rm eV\,nm^2$.}
\label{fig3}
\end{figure*}

Next, we perform numerical simulation of the
conductance oscillation on the lattice model.
Assuming that the size of both strip electrodes in the $\hat{z}$-direction
is much larger than the Fermi wavelength and their boundaries
are smooth enough, then the transverse momentum $k_z$ is approximately
conserved during scattering and can be regarded as a parameter. In this
way, the numerical calculation is reduced to a set of 2D slices labeled by $k_z$. For $|k_z|<k_0$,
a pair of edge states emerge under the open boundary condition [Fig. \ref{fig1}(c)].
By the substitutions $k_{i=x,y,z}\rightarrow\frac{1}{a}\sin k_ia$ and
$k_i^2\rightarrow\frac{2}{a^2}(1-\cos k_ia)$ while keeping $k_z$ as a parameter,
we obtain the lattice version of the Hamiltonian \eqref{H1} as
\begin{eqnarray}\label{HT}
\mathcal{H}_W(k_z)&=&\sum_{i} c_i^{\dagger}H_{ii}c_i+\sum_{i}(c_i^{\dagger}H_{i,i+a_x}c_{i+a_x}+H.c.)\nonumber\\
&+&\sum_{i}(c_i^{\dagger}H_{i,i+a_y}c_{i+a_y}+H.c.),
\end{eqnarray}
where $c_i=(c_{i,\uparrow},c_{i,\downarrow})$
is the Fermi operator on site $i=(i_x,i_y)$ with two pseudo-spin components,
$a_x=(a,0)$ and $a_y=(0,a)$ are the unit
vectors along the $\hat{x}$ and $\hat{y}$ direction, respectively,
with $a$ the lattice constant.
$H_{ii}$ and $H_{i,i+a_x(a_y)}$ are $2\times 2$ block matrices and
take the explicit forms as

\begin{figure*}
\centering
\includegraphics[width=\textwidth]{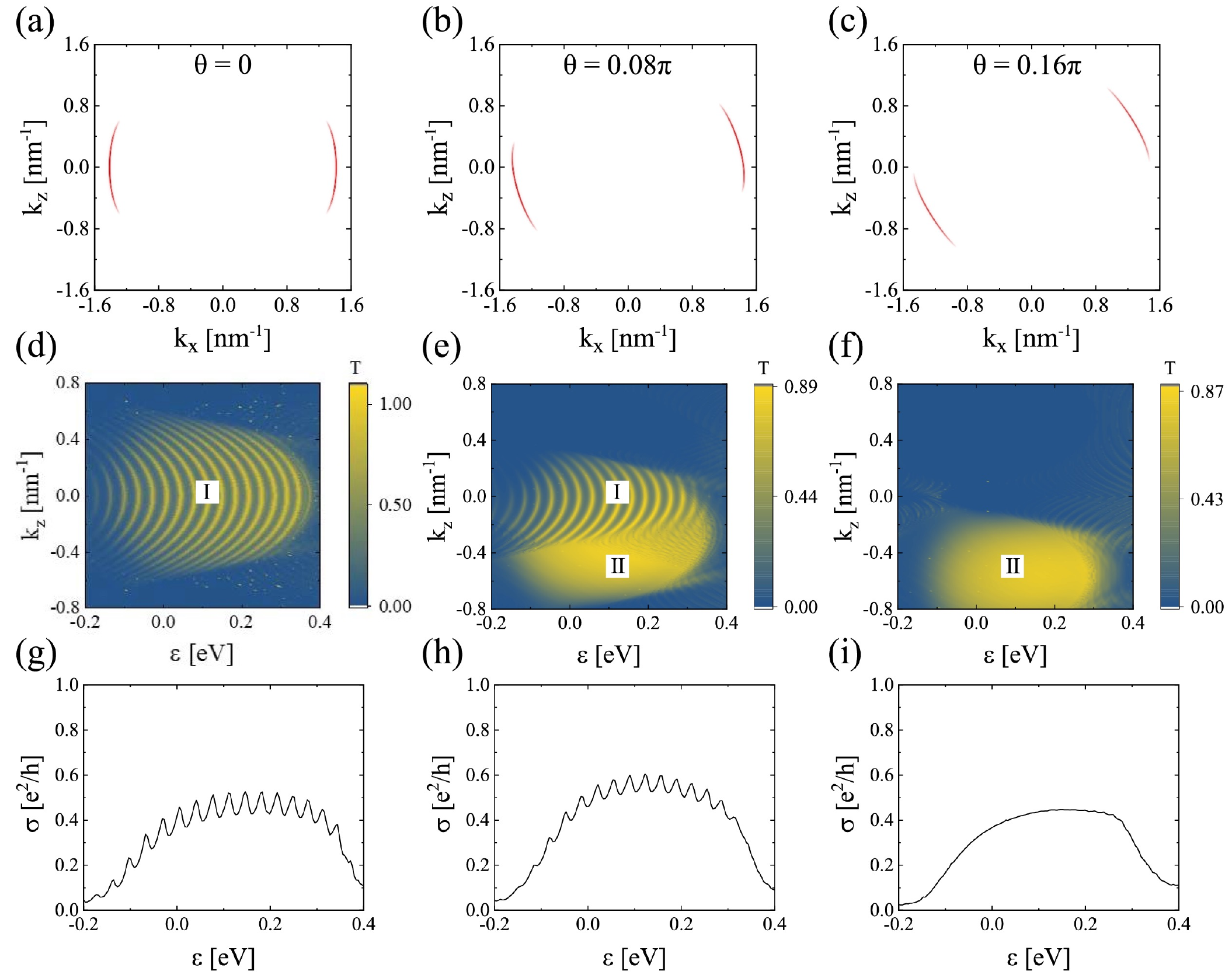}
\caption{(Color online).(a)-(c)Upper panel: FA spectra for different azimuthal angles $\theta$. (d)-(f)Middle panel: Corresponding transmission probability as a function of $\varepsilon$ and $k_z$. (g)-(i)Lower panel: Corresponding differential conductance after summing over all the transverse channels $k_z$. The other parameters are the same as those in Fig.\ref{fig3}(g).}
\label{fig4}
\end{figure*}

\begin{equation}\label{hop}
\begin{split}
H&_{ii}=M_1(k_1^2-\frac{2}{a^2})\sigma_x
+M_2(k_0^2-\frac{4}{a^2}+\frac{2}{a^2}\cos k_za)\sigma_z, \\
H&_{i,i+a_x}=\frac{M_1}{a^2}\sigma_x,\quad H_{i,i+a_y}=\frac{v_yi}{2a}\sigma_y+\frac{M_2}{a^2}\sigma_z.
\end{split}
\end{equation}
The configuration of FAs on the top surface can be revealed by the spectra function
$\mathcal{A}(\varepsilon)=-\frac{1}{\pi}\text{Im} g_W^R(\varepsilon)$ in the
top layer at $\varepsilon=0$,
with $g_W^R(\varepsilon,k_x,k_z)$ the retarded Green's function
calculated by the lattice model of the Weyl semimetal under open boundary condition
in the $\hat{y}$-direction; see Fig. \ref{fig3}(a).
To simulate the FAs in real materials \cite{PhysRevB.102.085126,PhysRevB.93.201101,belopolski2017signatures,PhysRevB.95.241108},
we have introduced an on-site potential $V$ on the top layer of the WSM lattice to introduce surface
dispersion that yields curved FAs \cite{PhysRevB.101.125407,zheng2020andreev}.

The strip electrodes can be described by an effective Hamiltonian
$H_N(\bm k)=(C\bm k^2-\mu_N)\sigma_0$, with $C$
the parameter related to the effective mass, $\mu_N$ the chemical
potential, and $\sigma_0$ the identity matrix.
The lattice model for the electrodes can be obtained in
a similar way as

\begin{equation}
\mathcal{H}_N(k_z)=\sum_jd_j^{\dagger}\lambda_j d_j-\frac{C}{a^2}\sum_j(d_j^{\dagger}d_{j+a_x}+d_j^{\dagger}d_{j+a_y}+H.c.),
\end{equation}
where $d_j=(d_{j,\uparrow},d_{j,\downarrow})$ is the Fermi operator, and $\lambda_j=2C(3-\cos k_za)/a^2-\mu_N$.

The whole system for a given $k_z$ is described by
$\mathcal{H}_W(k_z)$ and $\mathcal{H}_N(k_z)$ and the coupling between
them which is captured by the tunneling between the outmost lattice layers
with a strength $t_N$. The
thickness of the WSM and the electrodes in the
$\hat{y}$-direction is $100\,\rm nm$ and $40\,\rm nm$, respectively.
The width of the hopping area in the $\hat{x}$-direction is $W = 30\,\rm nm$,
and the separation between two
electrodes is $L =180\,\rm nm$ [cf. Fig. \ref{fig1}(a)].
Two on-site potentials $U_1$ and $U_2$ are introduced at
the boundary of the N electrodes to simulate the interface
barrier or the momentum mismatch in the heterostructure.
Both the WSM and electrodes connect to the leads extended to infinity
in the $\pm\hat{x}$ directions. The transmission
$T(k_z,\varepsilon)$ between two electrodes is calculated using
the KWANT program \cite{Groth_2014}. The overall conductance
by summing up all the transverse channels can be obtained as
\begin{equation}
\sigma(\varepsilon)=\frac{e^2}{h}\int_{-k_0}^{k_0}T(k_z,\varepsilon)dk_z.
\end{equation}
In the case of $\theta=0$, the FAs with different curvature and
length are shown in Figs. \ref{fig3}(a)-\ref{fig3}(c). The corresponding
results of the transmission probability $T(k_z,\varepsilon)$ and the
differential conductance $\sigma(\varepsilon)$ are plotted in Figs. \ref{fig3}(d)-\ref{fig3}(f)
and Figs. \ref{fig3}(g)-\ref{fig3}(i), respectively.
One can see from Fig. \ref{fig3} that less curved and shorter FAs
result in more visible conductance oscillation, in which the
dephasing effect between transverse channels becomes
weaker as revealed by the length and curvature of the
bright stripes in the transmission pattern in
Figs. \ref{fig3}(d)-\ref{fig3}(f). These results are
in coincidence with the analytical
calculations in Fig.\ref{fig2}.

\begin{figure*}
\centering
\includegraphics[width=\textwidth]{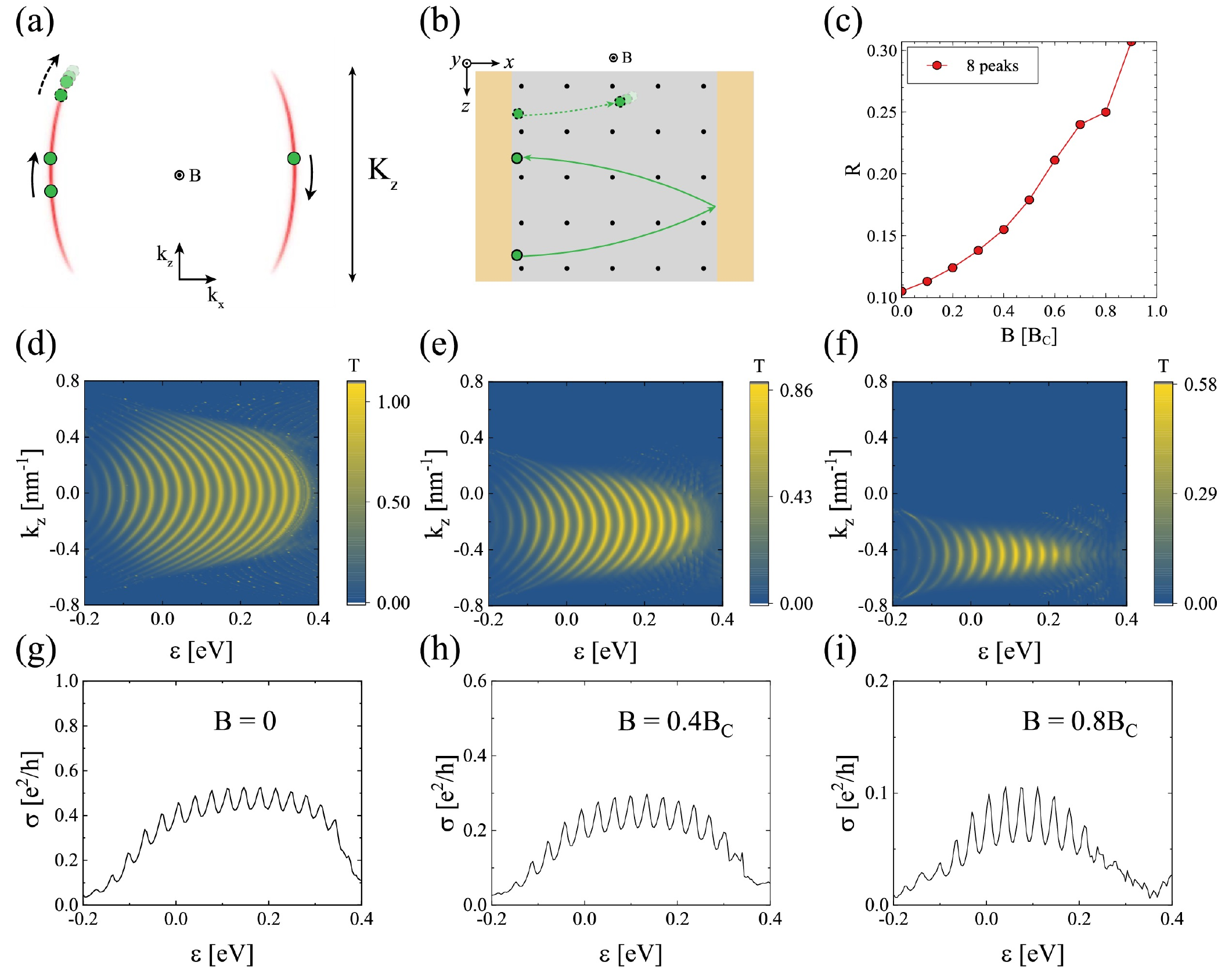}
\caption{(Color online).(a) FA spectrum for azimuthal angle
$\theta=0$, and electrons slide along the FA driven by the
Lorentz force. (b) Trajectories of electrons in real space
corresponding to the left panel. (c) Resolution of the conductance as a function
of the magnetic field, where $B_C$ is the saturated magnetic field. The transmission probability
as a function of $\varepsilon$ and $k_z$ with (d) $B=0$ (e) $B=0.4B_C$
(f) $B=0.8B_C$. (g)-(i)The corresponding conductance after summing over all
the transverse channels $k_z$ for different magnetic fields
in the $\hat{y}$ direction. The other parameters are the same
for those in Fig.\ref{fig3}(g).}
\label{fig5}
\end{figure*}

\section{Orientation dependent conductance spectra}\label{s4}
In the above section, we have seen that the shape
of the FAs strongly affect the oscillation pattern of the conductance.
In addition, FAs can also have diverse orientations
relative to the normal of the strip electrodes
with $\theta\neq 0$. Experimentally, this
can be achieved by fabricating the strip electrodes
along intended directions and theoretically, this
can be described by the effective model $H_W(\bm k,\theta)$ in Eq. \eqref{eq5}
with a finite rotation
while keeping the normal of the electrodes fixed to the $\hat{x}$-direction.

The numerical calculation is performed on the discretized version of
Hamiltonian $H_W(\bm k,\theta)$ in the same way as that in the previous section.
Again, the momentum $k_z$ is taken as a parameter. In
Figs. \ref{fig4}(a)-\ref{fig4}(c), we plot the FAs with different azimuthal
angles $\theta$. One can see that
the rotation of the effective Hamiltonian
causes corresponding rotated FAs.
The transmission probabilities $T(k_z,\varepsilon)$
as a function of energy $\varepsilon$ and
$k_z$ are shown in Figs. \ref{fig4}(d)-\ref{fig4}(f).
One can see that there in general exist two distinct regions
in the transmission pattern as shown in Fig. \ref{fig4}(e)
excepted for two limiting cases $\theta=0$ in Figs. \ref{fig4}(d) and
$\theta\geq\tan^{-1}(k_0/k_1)$ \ref{fig4}(f). Specifically,
the regions with stripe structures and nearly
uniform strength correspond to
region I and II in Fig. \ref{fig1}(b), respectively.
In region I, backscattering channels are available
which induces interference and oscillation of the transmission,
while in region II, the transport channels are chiral
with a high transmission without oscillation.
The amounts of $k_z$ channels in region I and II
vary with $\theta$, which is clearly
revealed in the conductance spectra in Figs. \ref{fig4}(g-i).
As $\theta$ increases from zero,
the oscillation of conductance becomes less
visible, stemming from that
the ratio between the numbers of $k_z$ channels
in region I and II becomes smaller.
When $\theta$ exceeds the threshold $\tan^{-1}(k_0/k_1)$
[Fig. \ref{fig4}(c)], all electrons reside in region $\rm{II}$
and the conductance exhibits a plateau structure without any oscillation as shown in
Fig. \ref{fig4}(i). Such a transition from oscillation to plateau
structure in the conductance spectra provides a clear manifestation of the highly
anisotropic nature of the FAs, and therefore can serve as its unique signal.
Although we elucidate such an effect
based on a specific model in Eq. \eqref{eq5},
the underlying physics should generally hold
for other WSMs with more complicated FA configurations.
The shape of FAs and the relative amount of $k_z$ channels
lying in region I and II are of most importance for the main results.

\section{magnetic field effect}\label{s5}
In this section, we show that the visibility of
the conductance oscillation can be
well improved by a magnetic field in the $\hat{y}$-direction.
We focus on the case $\theta=0$ and the results are
shown in Fig. \ref{fig5}.
The Landau gauge $\bm A=(0,0,-Bx)$ is adopted
such that the Peierls substitution
$\bm k\rightarrow -i\bm\nabla-e\bm{A}/\hbar$ (with $e>0$)
retains the $k_z$ conservation. For a small magnetic field
which satisfies
$B\ll \hbar k_{0,1}/(e a)$, it
only introduces a smooth modification
of the mass term in Eq. \eqref{H1}.
Such a pseudo-spin dependent potential
contributes an additional phase factor in the
transmission function $T(k_z,\varepsilon, B)$, which causes a shift of the pattern
in the $k_z$-direction as can be seen in Figs. \ref{fig5}(d-f).
A constant gauge term $\delta A$ can always
be added to the vector potential as $\bm A'=(0,0,-Bx+\delta A)$,
which is equivalent to an overall shift of $k_z$.
The physical results of the conductance spectra should not rely
on such freedom of gauge choice by noting
that the Hamiltonian is a periodic function of $k_z$,
so that an overall shift of $k_z$ by $\delta A$
has no effect after integration.
The magnetic field effect can also be well understood
by the semiclassical picture of
Lorentz force.
The Lorentz force drives electrons sliding along the FAs [Fig. \ref{fig5}(a)]
which corresponds to the curved trajectory in real space [Fig. \ref{fig5}(b)].

Remarkably, because that FAs are terminated at the Weyl points,
some of the electrons nearby can transfer into
the chiral Landau bands of the bulk states and dissipate due to the
surface-bulk connection at the Weyl points \cite{potter2014quantum} as illustrated by the dashed
green circles in Fig. \ref{fig5}(a)\cite{zheng2020andreev}.
As a result, these electrons can not reach the right electrode
and do not contribute to the conductance.
Therefore, the magnetic field effectively reduces
the number of $k_z$ channels and thus the dephasing
effect, which is reflected
in Figs. \ref{fig5}(d)-\ref{fig5}(f)
that the interference patterns get narrower in the $k_z$-direction as $B$ increases.
Accordingly, one can see in Figs. \ref{fig5}(g)-\ref{fig5}(i)
that as $B$ increases, the magnitude of the conductance reduces
due to the loss of surface electrons; However,
the oscillation of the conductance becomes more visible
stemming from weaker dephasing. To quantify the visibility
of conductance oscillation, we introduce the resolution defined as follows

\begin{equation}\label{res}
R=\sum_{i}^{8}\frac{\sigma_i^{\text{max}}-\sigma_i^{\text{min}}}{\sigma_i^{\text{max}}+\sigma_i^{\text{min}}},
\end{equation}
where $\sigma_i^{\text{max}}$ and $\sigma_i^{\text{min}}$
are the neighbouring maximum and minimum values of the conductance.
We choose the most visible eight oscillating periods
to calculate the resolution and plot $R$ as
a function of $B$ in Fig. \ref{fig5}(c). We see that
the resolution increases significantly with increasing magnetic field,
indicating that the observation of 2D Fabry-P\'{e}rot interference
can be facilitated by applying a magnetic field,
in stark contrast to other 2D electronic systems ~\cite{2008Quantum,2013Ballistic,2013A,
Oksanen2014Single,0Quantum,PhysRevLett.113.116601,PhysRevX.10.031007}.
Such a novel effect can also be used as
a direct evidence of FAs.

We remark that there exists a critical magnetic field $B_c=\hbar K_z/(eL)$
in the calculation,
above which all the incident
electrons in the FA surface states will transfer into the bulk
and no surface transport occurs. Here, $K_z$ is the span of the
FA in the $k_z$ direction [Fig. \ref{fig5}(a)] and $L$ is
the distance between the two electrodes [Fig. \ref{fig1}(a)].
For parameters $K_z=1.2\,\rm nm^{-1}$
and $L=180\,\rm nm$ adopted in Fig. \ref{fig5}, we have $B_c\simeq4.4$ Tesla.

\section{discussion and summary}\label{s6}

We would like to discuss the experimental realization
of our proposal. The surface NFAN junctions can
be achieved by state-of-the-art fabrication
techniques \cite{li2020fermi,chen2018finite,ghatak2018anomalous}.
The WSM with a pair of FAs is a crucial
building block in our proposal, which has
been reported in $\rm NbIrTe_4$ \cite{PhysRevB.102.085126,PhysRevB.93.201101,belopolski2017signatures,PhysRevB.95.241108},
$\rm WP_2$ \cite{PhysRevLett.122.176402}, $\rm MoTe_2$
\cite{PhysRevLett.117.056805}, and $\rm YbMnBi_2$ \cite{borisenko2019time}.
The main results still hold for other materials with more FAs
as long as the two regions of transmission can be well defined.
During the calculation, for simplicity, we set the chemical
potential being zero in the WSM to have a vanishing density
of the bulk states. In real materials with finite density of
states, our main conclusions remain unchanged as long as
the FAs and the bulk states are well separated in the
surface Brillouin zone. The presence of bulk states
will only cause
certain leakage of surface electrons, but will not
change the present qualitative results.

To summarize, we have investigated the 2D
conductance oscillation in the planar
NFAN junctions on the WSM
surface, which provides a unique transport signature
of the FAs. It is found that (i) Shorter and less curved FAs
can lead to more visible conductance oscillation;
(ii) A crossover from oscillation to plateau
structure of the conductance spectra
can be implemented by changing the orientation
of the planar junctions; (iii) The
magnetic field can significantly enhance the visibility
of the oscillation pattern which is unique for
the FA surface states.
Therefore, our work offers an effective way for the
identification of FA surface states through transport measurement.
It also introduces a new platform to realize interesting 2D conductance oscillation
induced by Fabry-P\'{e}rot-type interference.

\begin{acknowledgements}
This work was supported by the National
Natural Science Foundation of
China under Grant No. 12074172 (W.C.), the startup
grant at Nanjing University (W.C.), the State
Key Program for Basic Researches of China
under Grants No. 2017YFA0303203 (D.Y.X.)
and the Excellent Programme at Nanjing University.
\end{acknowledgements}


\begin{thebibliography}{65}%
\makeatletter
\providecommand \@ifxundefined [1]{%
 \@ifx{#1\undefined}
}%
\providecommand \@ifnum [1]{%
 \ifnum #1\expandafter \@firstoftwo
 \else \expandafter \@secondoftwo
 \fi
}%
\providecommand \@ifx [1]{%
 \ifx #1\expandafter \@firstoftwo
 \else \expandafter \@secondoftwo
 \fi
}%
\providecommand \natexlab [1]{#1}%
\providecommand \enquote  [1]{``#1''}%
\providecommand \bibnamefont  [1]{#1}%
\providecommand \bibfnamefont [1]{#1}%
\providecommand \citenamefont [1]{#1}%
\providecommand \href@noop [0]{\@secondoftwo}%
\providecommand \href [0]{\begingroup \@sanitize@url \@href}%
\providecommand \@href[1]{\@@startlink{#1}\@@href}%
\providecommand \@@href[1]{\endgroup#1\@@endlink}%
\providecommand \@sanitize@url [0]{\catcode `\\12\catcode `\$12\catcode
  `\&12\catcode `\#12\catcode `\^12\catcode `\_12\catcode `\%12\relax}%
\providecommand \@@startlink[1]{}%
\providecommand \@@endlink[0]{}%
\providecommand \url  [0]{\begingroup\@sanitize@url \@url }%
\providecommand \@url [1]{\endgroup\@href {#1}{\urlprefix }}%
\providecommand \urlprefix  [0]{URL }%
\providecommand \Eprint [0]{\href }%
\providecommand \doibase [0]{http://dx.doi.org/}%
\providecommand \selectlanguage [0]{\@gobble}%
\providecommand \bibinfo  [0]{\@secondoftwo}%
\providecommand \bibfield  [0]{\@secondoftwo}%
\providecommand \translation [1]{[#1]}%
\providecommand \BibitemOpen [0]{}%
\providecommand \bibitemStop [0]{}%
\providecommand \bibitemNoStop [0]{.\EOS\space}%
\providecommand \EOS [0]{\spacefactor3000\relax}%
\providecommand \BibitemShut  [1]{\csname bibitem#1\endcsname}%
\let\auto@bib@innerbib\@empty
\bibitem [{\citenamefont {Weyl}(1929)}]{weyl1929gravitation}%
  \BibitemOpen
  \bibfield  {author} {\bibinfo {author} {\bibfnamefont {H.}~\bibnamefont
  {Weyl}},\ }\href {https://www.ncbi.nlm.nih.gov/pmc/articles/PMC522457/}
  {\bibfield  {journal} {\bibinfo  {journal} {Proceedings of the National
  Academy of Sciences of the United States of America}\ }\textbf {\bibinfo
  {volume} {15}},\ \bibinfo {pages} {323} (\bibinfo {year} {1929})}\BibitemShut
  {NoStop}%
\bibitem [{\citenamefont {Armitage}\ \emph {et~al.}(2018)\citenamefont
  {Armitage}, \citenamefont {Mele},\ and\ \citenamefont
  {Vishwanath}}]{RevModPhys.90.015001}%
  \BibitemOpen
  \bibfield  {author} {\bibinfo {author} {\bibfnamefont {N.~P.}\ \bibnamefont
  {Armitage}}, \bibinfo {author} {\bibfnamefont {E.~J.}\ \bibnamefont {Mele}},
  \ and\ \bibinfo {author} {\bibfnamefont {A.}~\bibnamefont {Vishwanath}},\
  }\href {\doibase 10.1103/RevModPhys.90.015001} {\bibfield  {journal}
  {\bibinfo  {journal} {Rev. Mod. Phys.}\ }\textbf {\bibinfo {volume} {90}},\
  \bibinfo {pages} {015001} (\bibinfo {year} {2018})}\BibitemShut {NoStop}%
\bibitem [{\citenamefont {Kajita}(2016)}]{RevModPhys.88.030501}%
  \BibitemOpen
  \bibfield  {author} {\bibinfo {author} {\bibfnamefont {T.}~\bibnamefont
  {Kajita}},\ }\href {\doibase 10.1103/RevModPhys.88.030501} {\bibfield
  {journal} {\bibinfo  {journal} {Rev. Mod. Phys.}\ }\textbf {\bibinfo {volume}
  {88}},\ \bibinfo {pages} {030501} (\bibinfo {year} {2016})}\BibitemShut
  {NoStop}%
\bibitem [{\citenamefont {McDonald}(2016)}]{RevModPhys.88.030502}%
  \BibitemOpen
  \bibfield  {author} {\bibinfo {author} {\bibfnamefont {A.~B.}\ \bibnamefont
  {McDonald}},\ }\href {\doibase 10.1103/RevModPhys.88.030502} {\bibfield
  {journal} {\bibinfo  {journal} {Rev. Mod. Phys.}\ }\textbf {\bibinfo {volume}
  {88}},\ \bibinfo {pages} {030502} (\bibinfo {year} {2016})}\BibitemShut
  {NoStop}%
\bibitem [{\citenamefont {Wan}\ \emph {et~al.}(2011)\citenamefont {Wan},
  \citenamefont {Turner}, \citenamefont {Vishwanath},\ and\ \citenamefont
  {Savrasov}}]{PhysRevB.83.205101}%
  \BibitemOpen
  \bibfield  {author} {\bibinfo {author} {\bibfnamefont {X.}~\bibnamefont
  {Wan}}, \bibinfo {author} {\bibfnamefont {A.~M.}\ \bibnamefont {Turner}},
  \bibinfo {author} {\bibfnamefont {A.}~\bibnamefont {Vishwanath}}, \ and\
  \bibinfo {author} {\bibfnamefont {S.~Y.}\ \bibnamefont {Savrasov}},\ }\href
  {\doibase 10.1103/PhysRevB.83.205101} {\bibfield  {journal} {\bibinfo
  {journal} {Phys. Rev. B}\ }\textbf {\bibinfo {volume} {83}},\ \bibinfo
  {pages} {205101} (\bibinfo {year} {2011})}\BibitemShut {NoStop}%
\bibitem [{\citenamefont {Murakami}(2007)}]{Shuichi_Murakami_2007}%
  \BibitemOpen
  \bibfield  {author} {\bibinfo {author} {\bibfnamefont {S.}~\bibnamefont
  {Murakami}},\ }\href {\doibase 10.1088/1367-2630/9/9/356} {\bibfield
  {journal} {\bibinfo  {journal} {New Journal of Physics}\ }\textbf {\bibinfo
  {volume} {9}},\ \bibinfo {pages} {356} (\bibinfo {year} {2007})}\BibitemShut
  {NoStop}%
\bibitem [{\citenamefont {Burkov}\ and\ \citenamefont
  {Balents}(2011)}]{PhysRevLett.107.127205}%
  \BibitemOpen
  \bibfield  {author} {\bibinfo {author} {\bibfnamefont {A.~A.}\ \bibnamefont
  {Burkov}}\ and\ \bibinfo {author} {\bibfnamefont {L.}~\bibnamefont
  {Balents}},\ }\href {\doibase 10.1103/PhysRevLett.107.127205} {\bibfield
  {journal} {\bibinfo  {journal} {Phys. Rev. Lett.}\ }\textbf {\bibinfo
  {volume} {107}},\ \bibinfo {pages} {127205} (\bibinfo {year}
  {2011})}\BibitemShut {NoStop}%
\bibitem [{\citenamefont {Weng}\ \emph {et~al.}(2015)\citenamefont {Weng},
  \citenamefont {Fang}, \citenamefont {Fang}, \citenamefont {Bernevig},\ and\
  \citenamefont {Dai}}]{PhysRevX.5.011029}%
  \BibitemOpen
  \bibfield  {author} {\bibinfo {author} {\bibfnamefont {H.}~\bibnamefont
  {Weng}}, \bibinfo {author} {\bibfnamefont {C.}~\bibnamefont {Fang}}, \bibinfo
  {author} {\bibfnamefont {Z.}~\bibnamefont {Fang}}, \bibinfo {author}
  {\bibfnamefont {B.~A.}\ \bibnamefont {Bernevig}}, \ and\ \bibinfo {author}
  {\bibfnamefont {X.}~\bibnamefont {Dai}},\ }\href {\doibase
  10.1103/PhysRevX.5.011029} {\bibfield  {journal} {\bibinfo  {journal} {Phys.
  Rev. X}\ }\textbf {\bibinfo {volume} {5}},\ \bibinfo {pages} {011029}
  (\bibinfo {year} {2015})}\BibitemShut {NoStop}%
\bibitem [{\citenamefont {Huang}\ \emph
  {et~al.}(2015{\natexlab{a}})\citenamefont {Huang}, \citenamefont {Xu},
  \citenamefont {Belopolski}, \citenamefont {Lee}, \citenamefont {Chang},
  \citenamefont {Wang}, \citenamefont {Alidoust}, \citenamefont {Bian},
  \citenamefont {Neupane}, \citenamefont {Zhang} \emph
  {et~al.}}]{huang2015weyl}%
  \BibitemOpen
  \bibfield  {author} {\bibinfo {author} {\bibfnamefont {S.-M.}\ \bibnamefont
  {Huang}}, \bibinfo {author} {\bibfnamefont {S.-Y.}\ \bibnamefont {Xu}},
  \bibinfo {author} {\bibfnamefont {I.}~\bibnamefont {Belopolski}}, \bibinfo
  {author} {\bibfnamefont {C.-C.}\ \bibnamefont {Lee}}, \bibinfo {author}
  {\bibfnamefont {G.}~\bibnamefont {Chang}}, \bibinfo {author} {\bibfnamefont
  {B.}~\bibnamefont {Wang}}, \bibinfo {author} {\bibfnamefont {N.}~\bibnamefont
  {Alidoust}}, \bibinfo {author} {\bibfnamefont {G.}~\bibnamefont {Bian}},
  \bibinfo {author} {\bibfnamefont {M.}~\bibnamefont {Neupane}}, \bibinfo
  {author} {\bibfnamefont {C.}~\bibnamefont {Zhang}},  \emph {et~al.},\ }\href
  {https://www.nature.com/articles/ncomms8373} {\bibfield  {journal} {\bibinfo
  {journal} {Nature communications}\ }\textbf {\bibinfo {volume} {6}},\
  \bibinfo {pages} {1} (\bibinfo {year} {2015}{\natexlab{a}})}\BibitemShut
  {NoStop}%
\bibitem [{\citenamefont {Lv}\ \emph {et~al.}(2015{\natexlab{a}})\citenamefont
  {Lv}, \citenamefont {Weng}, \citenamefont {Fu}, \citenamefont {Wang},
  \citenamefont {Miao}, \citenamefont {Ma}, \citenamefont {Richard},
  \citenamefont {Huang}, \citenamefont {Zhao}, \citenamefont {Chen},
  \citenamefont {Fang}, \citenamefont {Dai}, \citenamefont {Qian},\ and\
  \citenamefont {Ding}}]{PhysRevX.5.031013}%
  \BibitemOpen
  \bibfield  {author} {\bibinfo {author} {\bibfnamefont {B.~Q.}\ \bibnamefont
  {Lv}}, \bibinfo {author} {\bibfnamefont {H.~M.}\ \bibnamefont {Weng}},
  \bibinfo {author} {\bibfnamefont {B.~B.}\ \bibnamefont {Fu}}, \bibinfo
  {author} {\bibfnamefont {X.~P.}\ \bibnamefont {Wang}}, \bibinfo {author}
  {\bibfnamefont {H.}~\bibnamefont {Miao}}, \bibinfo {author} {\bibfnamefont
  {J.}~\bibnamefont {Ma}}, \bibinfo {author} {\bibfnamefont {P.}~\bibnamefont
  {Richard}}, \bibinfo {author} {\bibfnamefont {X.~C.}\ \bibnamefont {Huang}},
  \bibinfo {author} {\bibfnamefont {L.~X.}\ \bibnamefont {Zhao}}, \bibinfo
  {author} {\bibfnamefont {G.~F.}\ \bibnamefont {Chen}}, \bibinfo {author}
  {\bibfnamefont {Z.}~\bibnamefont {Fang}}, \bibinfo {author} {\bibfnamefont
  {X.}~\bibnamefont {Dai}}, \bibinfo {author} {\bibfnamefont {T.}~\bibnamefont
  {Qian}}, \ and\ \bibinfo {author} {\bibfnamefont {H.}~\bibnamefont {Ding}},\
  }\href {\doibase 10.1103/PhysRevX.5.031013} {\bibfield  {journal} {\bibinfo
  {journal} {Phys. Rev. X}\ }\textbf {\bibinfo {volume} {5}},\ \bibinfo {pages}
  {031013} (\bibinfo {year} {2015}{\natexlab{a}})}\BibitemShut {NoStop}%
\bibitem [{\citenamefont {Xu}\ \emph {et~al.}(2015{\natexlab{a}})\citenamefont
  {Xu}, \citenamefont {Belopolski}, \citenamefont {Alidoust}, \citenamefont
  {Neupane}, \citenamefont {Bian}, \citenamefont {Zhang}, \citenamefont
  {Sankar}, \citenamefont {Chang}, \citenamefont {Yuan}, \citenamefont {Lee},
  \citenamefont {Huang}, \citenamefont {Zheng}, \citenamefont {Ma},
  \citenamefont {Sanchez}, \citenamefont {Wang}, \citenamefont {Bansil},
  \citenamefont {Chou}, \citenamefont {Shibayev}, \citenamefont {Lin},
  \citenamefont {Jia},\ and\ \citenamefont {Hasan}}]{Xu613}%
  \BibitemOpen
  \bibfield  {author} {\bibinfo {author} {\bibfnamefont {S.-Y.}\ \bibnamefont
  {Xu}}, \bibinfo {author} {\bibfnamefont {I.}~\bibnamefont {Belopolski}},
  \bibinfo {author} {\bibfnamefont {N.}~\bibnamefont {Alidoust}}, \bibinfo
  {author} {\bibfnamefont {M.}~\bibnamefont {Neupane}}, \bibinfo {author}
  {\bibfnamefont {G.}~\bibnamefont {Bian}}, \bibinfo {author} {\bibfnamefont
  {C.}~\bibnamefont {Zhang}}, \bibinfo {author} {\bibfnamefont
  {R.}~\bibnamefont {Sankar}}, \bibinfo {author} {\bibfnamefont
  {G.}~\bibnamefont {Chang}}, \bibinfo {author} {\bibfnamefont
  {Z.}~\bibnamefont {Yuan}}, \bibinfo {author} {\bibfnamefont {C.-C.}\
  \bibnamefont {Lee}}, \bibinfo {author} {\bibfnamefont {S.-M.}\ \bibnamefont
  {Huang}}, \bibinfo {author} {\bibfnamefont {H.}~\bibnamefont {Zheng}},
  \bibinfo {author} {\bibfnamefont {J.}~\bibnamefont {Ma}}, \bibinfo {author}
  {\bibfnamefont {D.~S.}\ \bibnamefont {Sanchez}}, \bibinfo {author}
  {\bibfnamefont {B.}~\bibnamefont {Wang}}, \bibinfo {author} {\bibfnamefont
  {A.}~\bibnamefont {Bansil}}, \bibinfo {author} {\bibfnamefont
  {F.}~\bibnamefont {Chou}}, \bibinfo {author} {\bibfnamefont {P.~P.}\
  \bibnamefont {Shibayev}}, \bibinfo {author} {\bibfnamefont {H.}~\bibnamefont
  {Lin}}, \bibinfo {author} {\bibfnamefont {S.}~\bibnamefont {Jia}}, \ and\
  \bibinfo {author} {\bibfnamefont {M.~Z.}\ \bibnamefont {Hasan}},\ }\href
  {\doibase 10.1126/science.aaa9297} {\bibfield  {journal} {\bibinfo  {journal}
  {Science}\ }\textbf {\bibinfo {volume} {349}},\ \bibinfo {pages} {613}
  (\bibinfo {year} {2015}{\natexlab{a}})}\BibitemShut {NoStop}%
\bibitem [{\citenamefont {Xu}\ \emph {et~al.}(2015{\natexlab{b}})\citenamefont
  {Xu}, \citenamefont {Alidoust}, \citenamefont {Belopolski}, \citenamefont
  {Yuan}, \citenamefont {Bian}, \citenamefont {Chang}, \citenamefont {Zheng},
  \citenamefont {Strocov}, \citenamefont {Sanchez}, \citenamefont {Chang} \emph
  {et~al.}}]{xu2015discovery}%
  \BibitemOpen
  \bibfield  {author} {\bibinfo {author} {\bibfnamefont {S.-Y.}\ \bibnamefont
  {Xu}}, \bibinfo {author} {\bibfnamefont {N.}~\bibnamefont {Alidoust}},
  \bibinfo {author} {\bibfnamefont {I.}~\bibnamefont {Belopolski}}, \bibinfo
  {author} {\bibfnamefont {Z.}~\bibnamefont {Yuan}}, \bibinfo {author}
  {\bibfnamefont {G.}~\bibnamefont {Bian}}, \bibinfo {author} {\bibfnamefont
  {T.-R.}\ \bibnamefont {Chang}}, \bibinfo {author} {\bibfnamefont
  {H.}~\bibnamefont {Zheng}}, \bibinfo {author} {\bibfnamefont {V.~N.}\
  \bibnamefont {Strocov}}, \bibinfo {author} {\bibfnamefont {D.~S.}\
  \bibnamefont {Sanchez}}, \bibinfo {author} {\bibfnamefont {G.}~\bibnamefont
  {Chang}},  \emph {et~al.},\ }\href
  {https://www.nature.com/articles/nphys3437} {\bibfield  {journal} {\bibinfo
  {journal} {Nature Physics}\ }\textbf {\bibinfo {volume} {11}},\ \bibinfo
  {pages} {748} (\bibinfo {year} {2015}{\natexlab{b}})}\BibitemShut {NoStop}%
\bibitem [{\citenamefont {Xu}\ \emph {et~al.}(2015{\natexlab{c}})\citenamefont
  {Xu}, \citenamefont {Belopolski}, \citenamefont {Sanchez}, \citenamefont
  {Zhang}, \citenamefont {Chang}, \citenamefont {Guo}, \citenamefont {Bian},
  \citenamefont {Yuan}, \citenamefont {Lu}, \citenamefont {Chang},
  \citenamefont {Shibayev}, \citenamefont {Prokopovych}, \citenamefont
  {Alidoust}, \citenamefont {Zheng}, \citenamefont {Lee}, \citenamefont
  {Huang}, \citenamefont {Sankar}, \citenamefont {Chou}, \citenamefont {Hsu},
  \citenamefont {Jeng}, \citenamefont {Bansil}, \citenamefont {Neupert},
  \citenamefont {Strocov}, \citenamefont {Lin}, \citenamefont {Jia},\ and\
  \citenamefont {Hasan}}]{Xue1501092}%
  \BibitemOpen
  \bibfield  {author} {\bibinfo {author} {\bibfnamefont {S.-Y.}\ \bibnamefont
  {Xu}}, \bibinfo {author} {\bibfnamefont {I.}~\bibnamefont {Belopolski}},
  \bibinfo {author} {\bibfnamefont {D.~S.}\ \bibnamefont {Sanchez}}, \bibinfo
  {author} {\bibfnamefont {C.}~\bibnamefont {Zhang}}, \bibinfo {author}
  {\bibfnamefont {G.}~\bibnamefont {Chang}}, \bibinfo {author} {\bibfnamefont
  {C.}~\bibnamefont {Guo}}, \bibinfo {author} {\bibfnamefont {G.}~\bibnamefont
  {Bian}}, \bibinfo {author} {\bibfnamefont {Z.}~\bibnamefont {Yuan}}, \bibinfo
  {author} {\bibfnamefont {H.}~\bibnamefont {Lu}}, \bibinfo {author}
  {\bibfnamefont {T.-R.}\ \bibnamefont {Chang}}, \bibinfo {author}
  {\bibfnamefont {P.~P.}\ \bibnamefont {Shibayev}}, \bibinfo {author}
  {\bibfnamefont {M.~L.}\ \bibnamefont {Prokopovych}}, \bibinfo {author}
  {\bibfnamefont {N.}~\bibnamefont {Alidoust}}, \bibinfo {author}
  {\bibfnamefont {H.}~\bibnamefont {Zheng}}, \bibinfo {author} {\bibfnamefont
  {C.-C.}\ \bibnamefont {Lee}}, \bibinfo {author} {\bibfnamefont {S.-M.}\
  \bibnamefont {Huang}}, \bibinfo {author} {\bibfnamefont {R.}~\bibnamefont
  {Sankar}}, \bibinfo {author} {\bibfnamefont {F.}~\bibnamefont {Chou}},
  \bibinfo {author} {\bibfnamefont {C.-H.}\ \bibnamefont {Hsu}}, \bibinfo
  {author} {\bibfnamefont {H.-T.}\ \bibnamefont {Jeng}}, \bibinfo {author}
  {\bibfnamefont {A.}~\bibnamefont {Bansil}}, \bibinfo {author} {\bibfnamefont
  {T.}~\bibnamefont {Neupert}}, \bibinfo {author} {\bibfnamefont {V.~N.}\
  \bibnamefont {Strocov}}, \bibinfo {author} {\bibfnamefont {H.}~\bibnamefont
  {Lin}}, \bibinfo {author} {\bibfnamefont {S.}~\bibnamefont {Jia}}, \ and\
  \bibinfo {author} {\bibfnamefont {M.~Z.}\ \bibnamefont {Hasan}},\ }\href
  {\doibase 10.1126/sciadv.1501092} {\bibfield  {journal} {\bibinfo  {journal}
  {Science Advances}\ }\textbf {\bibinfo {volume} {1}} (\bibinfo {year}
  {2015}{\natexlab{c}}),\ 10.1126/sciadv.1501092}\BibitemShut {NoStop}%
\bibitem [{\citenamefont {Xu}\ \emph {et~al.}(2016)\citenamefont {Xu},
  \citenamefont {Weng}, \citenamefont {Lv}, \citenamefont {Matt}, \citenamefont
  {Park}, \citenamefont {Bisti}, \citenamefont {Strocov}, \citenamefont
  {Gawryluk}, \citenamefont {Pomjakushina}, \citenamefont {Conder} \emph
  {et~al.}}]{xu2016observation}%
  \BibitemOpen
  \bibfield  {author} {\bibinfo {author} {\bibfnamefont {N.}~\bibnamefont
  {Xu}}, \bibinfo {author} {\bibfnamefont {H.}~\bibnamefont {Weng}}, \bibinfo
  {author} {\bibfnamefont {B.}~\bibnamefont {Lv}}, \bibinfo {author}
  {\bibfnamefont {C.~E.}\ \bibnamefont {Matt}}, \bibinfo {author}
  {\bibfnamefont {J.}~\bibnamefont {Park}}, \bibinfo {author} {\bibfnamefont
  {F.}~\bibnamefont {Bisti}}, \bibinfo {author} {\bibfnamefont {V.~N.}\
  \bibnamefont {Strocov}}, \bibinfo {author} {\bibfnamefont {D.}~\bibnamefont
  {Gawryluk}}, \bibinfo {author} {\bibfnamefont {E.}~\bibnamefont
  {Pomjakushina}}, \bibinfo {author} {\bibfnamefont {K.}~\bibnamefont
  {Conder}},  \emph {et~al.},\ }\href
  {https://www.nature.com/articles/ncomms11006} {\bibfield  {journal} {\bibinfo
   {journal} {Nature communications}\ }\textbf {\bibinfo {volume} {7}},\
  \bibinfo {pages} {1} (\bibinfo {year} {2016})}\BibitemShut {NoStop}%
\bibitem [{\citenamefont {Deng}\ \emph {et~al.}(2016)\citenamefont {Deng},
  \citenamefont {Wan}, \citenamefont {Deng}, \citenamefont {Zhang},
  \citenamefont {Ding}, \citenamefont {Wang}, \citenamefont {Yan},
  \citenamefont {Huang}, \citenamefont {Zhang}, \citenamefont {Xu} \emph
  {et~al.}}]{deng2016experimental}%
  \BibitemOpen
  \bibfield  {author} {\bibinfo {author} {\bibfnamefont {K.}~\bibnamefont
  {Deng}}, \bibinfo {author} {\bibfnamefont {G.}~\bibnamefont {Wan}}, \bibinfo
  {author} {\bibfnamefont {P.}~\bibnamefont {Deng}}, \bibinfo {author}
  {\bibfnamefont {K.}~\bibnamefont {Zhang}}, \bibinfo {author} {\bibfnamefont
  {S.}~\bibnamefont {Ding}}, \bibinfo {author} {\bibfnamefont {E.}~\bibnamefont
  {Wang}}, \bibinfo {author} {\bibfnamefont {M.}~\bibnamefont {Yan}}, \bibinfo
  {author} {\bibfnamefont {H.}~\bibnamefont {Huang}}, \bibinfo {author}
  {\bibfnamefont {H.}~\bibnamefont {Zhang}}, \bibinfo {author} {\bibfnamefont
  {Z.}~\bibnamefont {Xu}},  \emph {et~al.},\ }\href
  {https://www.nature.com/articles/nphys3871} {\bibfield  {journal} {\bibinfo
  {journal} {Nature Physics}\ }\textbf {\bibinfo {volume} {12}},\ \bibinfo
  {pages} {1105} (\bibinfo {year} {2016})}\BibitemShut {NoStop}%
\bibitem [{\citenamefont {Yang}\ \emph {et~al.}(2015)\citenamefont {Yang},
  \citenamefont {Liu}, \citenamefont {Sun}, \citenamefont {Peng}, \citenamefont
  {Yang}, \citenamefont {Zhang}, \citenamefont {Zhou}, \citenamefont {Zhang},
  \citenamefont {Guo}, \citenamefont {Rahn} \emph {et~al.}}]{yang2015weyl}%
  \BibitemOpen
  \bibfield  {author} {\bibinfo {author} {\bibfnamefont {L.}~\bibnamefont
  {Yang}}, \bibinfo {author} {\bibfnamefont {Z.}~\bibnamefont {Liu}}, \bibinfo
  {author} {\bibfnamefont {Y.}~\bibnamefont {Sun}}, \bibinfo {author}
  {\bibfnamefont {H.}~\bibnamefont {Peng}}, \bibinfo {author} {\bibfnamefont
  {H.}~\bibnamefont {Yang}}, \bibinfo {author} {\bibfnamefont {T.}~\bibnamefont
  {Zhang}}, \bibinfo {author} {\bibfnamefont {B.}~\bibnamefont {Zhou}},
  \bibinfo {author} {\bibfnamefont {Y.}~\bibnamefont {Zhang}}, \bibinfo
  {author} {\bibfnamefont {Y.}~\bibnamefont {Guo}}, \bibinfo {author}
  {\bibfnamefont {M.}~\bibnamefont {Rahn}},  \emph {et~al.},\ }\href
  {https://www.nature.com/articles/nphys3425} {\bibfield  {journal} {\bibinfo
  {journal} {Nature physics}\ }\textbf {\bibinfo {volume} {11}},\ \bibinfo
  {pages} {728} (\bibinfo {year} {2015})}\BibitemShut {NoStop}%
\bibitem [{\citenamefont {Huang}\ \emph {et~al.}(2016)\citenamefont {Huang},
  \citenamefont {McCormick}, \citenamefont {Ochi}, \citenamefont {Zhao},
  \citenamefont {Suzuki}, \citenamefont {Arita}, \citenamefont {Wu},
  \citenamefont {Mou}, \citenamefont {Cao}, \citenamefont {Yan} \emph
  {et~al.}}]{huang2016spectroscopic}%
  \BibitemOpen
  \bibfield  {author} {\bibinfo {author} {\bibfnamefont {L.}~\bibnamefont
  {Huang}}, \bibinfo {author} {\bibfnamefont {T.~M.}\ \bibnamefont
  {McCormick}}, \bibinfo {author} {\bibfnamefont {M.}~\bibnamefont {Ochi}},
  \bibinfo {author} {\bibfnamefont {Z.}~\bibnamefont {Zhao}}, \bibinfo {author}
  {\bibfnamefont {M.-T.}\ \bibnamefont {Suzuki}}, \bibinfo {author}
  {\bibfnamefont {R.}~\bibnamefont {Arita}}, \bibinfo {author} {\bibfnamefont
  {Y.}~\bibnamefont {Wu}}, \bibinfo {author} {\bibfnamefont {D.}~\bibnamefont
  {Mou}}, \bibinfo {author} {\bibfnamefont {H.}~\bibnamefont {Cao}}, \bibinfo
  {author} {\bibfnamefont {J.}~\bibnamefont {Yan}},  \emph {et~al.},\ }\href
  {https://www.nature.com/articles/nmat4685} {\bibfield  {journal} {\bibinfo
  {journal} {Nature materials}\ }\textbf {\bibinfo {volume} {15}},\ \bibinfo
  {pages} {1155} (\bibinfo {year} {2016})}\BibitemShut {NoStop}%
\bibitem [{\citenamefont {Tamai}\ \emph {et~al.}(2016)\citenamefont {Tamai},
  \citenamefont {Wu}, \citenamefont {Cucchi}, \citenamefont {Bruno},
  \citenamefont {Ricc\`o}, \citenamefont {Kim}, \citenamefont {Hoesch},
  \citenamefont {Barreteau}, \citenamefont {Giannini}, \citenamefont {Besnard},
  \citenamefont {Soluyanov},\ and\ \citenamefont
  {Baumberger}}]{PhysRevX.6.031021}%
  \BibitemOpen
  \bibfield  {author} {\bibinfo {author} {\bibfnamefont {A.}~\bibnamefont
  {Tamai}}, \bibinfo {author} {\bibfnamefont {Q.~S.}\ \bibnamefont {Wu}},
  \bibinfo {author} {\bibfnamefont {I.}~\bibnamefont {Cucchi}}, \bibinfo
  {author} {\bibfnamefont {F.~Y.}\ \bibnamefont {Bruno}}, \bibinfo {author}
  {\bibfnamefont {S.}~\bibnamefont {Ricc\`o}}, \bibinfo {author} {\bibfnamefont
  {T.~K.}\ \bibnamefont {Kim}}, \bibinfo {author} {\bibfnamefont
  {M.}~\bibnamefont {Hoesch}}, \bibinfo {author} {\bibfnamefont
  {C.}~\bibnamefont {Barreteau}}, \bibinfo {author} {\bibfnamefont
  {E.}~\bibnamefont {Giannini}}, \bibinfo {author} {\bibfnamefont
  {C.}~\bibnamefont {Besnard}}, \bibinfo {author} {\bibfnamefont {A.~A.}\
  \bibnamefont {Soluyanov}}, \ and\ \bibinfo {author} {\bibfnamefont
  {F.}~\bibnamefont {Baumberger}},\ }\href {\doibase 10.1103/PhysRevX.6.031021}
  {\bibfield  {journal} {\bibinfo  {journal} {Phys. Rev. X}\ }\textbf {\bibinfo
  {volume} {6}},\ \bibinfo {pages} {031021} (\bibinfo {year}
  {2016})}\BibitemShut {NoStop}%
\bibitem [{\citenamefont {Jiang}\ \emph {et~al.}(2017)\citenamefont {Jiang},
  \citenamefont {Liu}, \citenamefont {Sun}, \citenamefont {Yang}, \citenamefont
  {Rajamathi}, \citenamefont {Qi}, \citenamefont {Yang}, \citenamefont {Chen},
  \citenamefont {Peng}, \citenamefont {Hwang} \emph
  {et~al.}}]{jiang2017signature}%
  \BibitemOpen
  \bibfield  {author} {\bibinfo {author} {\bibfnamefont {J.}~\bibnamefont
  {Jiang}}, \bibinfo {author} {\bibfnamefont {Z.}~\bibnamefont {Liu}}, \bibinfo
  {author} {\bibfnamefont {Y.}~\bibnamefont {Sun}}, \bibinfo {author}
  {\bibfnamefont {H.}~\bibnamefont {Yang}}, \bibinfo {author} {\bibfnamefont
  {C.}~\bibnamefont {Rajamathi}}, \bibinfo {author} {\bibfnamefont
  {Y.}~\bibnamefont {Qi}}, \bibinfo {author} {\bibfnamefont {L.}~\bibnamefont
  {Yang}}, \bibinfo {author} {\bibfnamefont {C.}~\bibnamefont {Chen}}, \bibinfo
  {author} {\bibfnamefont {H.}~\bibnamefont {Peng}}, \bibinfo {author}
  {\bibfnamefont {C.}~\bibnamefont {Hwang}},  \emph {et~al.},\ }\href
  {https://www.nature.com/articles/ncomms13973} {\bibfield  {journal} {\bibinfo
   {journal} {Nature communications}\ }\textbf {\bibinfo {volume} {8}},\
  \bibinfo {pages} {1} (\bibinfo {year} {2017})}\BibitemShut {NoStop}%
\bibitem [{\citenamefont {Belopolski}\ \emph {et~al.}(2016)\citenamefont
  {Belopolski}, \citenamefont {Sanchez}, \citenamefont {Ishida}, \citenamefont
  {Pan}, \citenamefont {Yu}, \citenamefont {Xu}, \citenamefont {Chang},
  \citenamefont {Chang}, \citenamefont {Zheng}, \citenamefont {Alidoust} \emph
  {et~al.}}]{belopolski2016discovery}%
  \BibitemOpen
  \bibfield  {author} {\bibinfo {author} {\bibfnamefont {I.}~\bibnamefont
  {Belopolski}}, \bibinfo {author} {\bibfnamefont {D.~S.}\ \bibnamefont
  {Sanchez}}, \bibinfo {author} {\bibfnamefont {Y.}~\bibnamefont {Ishida}},
  \bibinfo {author} {\bibfnamefont {X.}~\bibnamefont {Pan}}, \bibinfo {author}
  {\bibfnamefont {P.}~\bibnamefont {Yu}}, \bibinfo {author} {\bibfnamefont
  {S.-Y.}\ \bibnamefont {Xu}}, \bibinfo {author} {\bibfnamefont
  {G.}~\bibnamefont {Chang}}, \bibinfo {author} {\bibfnamefont {T.-R.}\
  \bibnamefont {Chang}}, \bibinfo {author} {\bibfnamefont {H.}~\bibnamefont
  {Zheng}}, \bibinfo {author} {\bibfnamefont {N.}~\bibnamefont {Alidoust}},
  \emph {et~al.},\ }\href {https://www.nature.com/articles/ncomms13643}
  {\bibfield  {journal} {\bibinfo  {journal} {Nature communications}\ }\textbf
  {\bibinfo {volume} {7}},\ \bibinfo {pages} {1} (\bibinfo {year}
  {2016})}\BibitemShut {NoStop}%
\bibitem [{\citenamefont {Lv}\ \emph {et~al.}(2015{\natexlab{b}})\citenamefont
  {Lv}, \citenamefont {Xu}, \citenamefont {Weng}, \citenamefont {Ma},
  \citenamefont {Richard}, \citenamefont {Huang}, \citenamefont {Zhao},
  \citenamefont {Chen}, \citenamefont {Matt}, \citenamefont {Bisti} \emph
  {et~al.}}]{lv2015observation}%
  \BibitemOpen
  \bibfield  {author} {\bibinfo {author} {\bibfnamefont {B.}~\bibnamefont
  {Lv}}, \bibinfo {author} {\bibfnamefont {N.}~\bibnamefont {Xu}}, \bibinfo
  {author} {\bibfnamefont {H.}~\bibnamefont {Weng}}, \bibinfo {author}
  {\bibfnamefont {J.}~\bibnamefont {Ma}}, \bibinfo {author} {\bibfnamefont
  {P.}~\bibnamefont {Richard}}, \bibinfo {author} {\bibfnamefont
  {X.}~\bibnamefont {Huang}}, \bibinfo {author} {\bibfnamefont
  {L.}~\bibnamefont {Zhao}}, \bibinfo {author} {\bibfnamefont {G.}~\bibnamefont
  {Chen}}, \bibinfo {author} {\bibfnamefont {C.}~\bibnamefont {Matt}}, \bibinfo
  {author} {\bibfnamefont {F.}~\bibnamefont {Bisti}},  \emph {et~al.},\ }\href
  {https://www.nature.com/articles/nphys3426} {\bibfield  {journal} {\bibinfo
  {journal} {Nature Physics}\ }\textbf {\bibinfo {volume} {11}},\ \bibinfo
  {pages} {724} (\bibinfo {year} {2015}{\natexlab{b}})}\BibitemShut {NoStop}%
\bibitem [{\citenamefont {Zyuzin}\ and\ \citenamefont
  {Burkov}(2012)}]{PhysRevB.86.115133}%
  \BibitemOpen
  \bibfield  {author} {\bibinfo {author} {\bibfnamefont {A.~A.}\ \bibnamefont
  {Zyuzin}}\ and\ \bibinfo {author} {\bibfnamefont {A.~A.}\ \bibnamefont
  {Burkov}},\ }\href {\doibase 10.1103/PhysRevB.86.115133} {\bibfield
  {journal} {\bibinfo  {journal} {Phys. Rev. B}\ }\textbf {\bibinfo {volume}
  {86}},\ \bibinfo {pages} {115133} (\bibinfo {year} {2012})}\BibitemShut
  {NoStop}%
\bibitem [{\citenamefont {Aji}(2012)}]{PhysRevB.85.241101}%
  \BibitemOpen
  \bibfield  {author} {\bibinfo {author} {\bibfnamefont {V.}~\bibnamefont
  {Aji}},\ }\href {\doibase 10.1103/PhysRevB.85.241101} {\bibfield  {journal}
  {\bibinfo  {journal} {Phys. Rev. B}\ }\textbf {\bibinfo {volume} {85}},\
  \bibinfo {pages} {241101} (\bibinfo {year} {2012})}\BibitemShut {NoStop}%
\bibitem [{\citenamefont {Son}\ and\ \citenamefont
  {Spivak}(2013)}]{PhysRevB.88.104412}%
  \BibitemOpen
  \bibfield  {author} {\bibinfo {author} {\bibfnamefont {D.~T.}\ \bibnamefont
  {Son}}\ and\ \bibinfo {author} {\bibfnamefont {B.~Z.}\ \bibnamefont
  {Spivak}},\ }\href {\doibase 10.1103/PhysRevB.88.104412} {\bibfield
  {journal} {\bibinfo  {journal} {Phys. Rev. B}\ }\textbf {\bibinfo {volume}
  {88}},\ \bibinfo {pages} {104412} (\bibinfo {year} {2013})}\BibitemShut
  {NoStop}%
\bibitem [{\citenamefont {Chernodub}\ \emph {et~al.}(2014)\citenamefont
  {Chernodub}, \citenamefont {Cortijo}, \citenamefont {Grushin}, \citenamefont
  {Landsteiner},\ and\ \citenamefont {Vozmediano}}]{PhysRevB.89.081407}%
  \BibitemOpen
  \bibfield  {author} {\bibinfo {author} {\bibfnamefont {M.~N.}\ \bibnamefont
  {Chernodub}}, \bibinfo {author} {\bibfnamefont {A.}~\bibnamefont {Cortijo}},
  \bibinfo {author} {\bibfnamefont {A.~G.}\ \bibnamefont {Grushin}}, \bibinfo
  {author} {\bibfnamefont {K.}~\bibnamefont {Landsteiner}}, \ and\ \bibinfo
  {author} {\bibfnamefont {M.~A.~H.}\ \bibnamefont {Vozmediano}},\ }\href
  {\doibase 10.1103/PhysRevB.89.081407} {\bibfield  {journal} {\bibinfo
  {journal} {Phys. Rev. B}\ }\textbf {\bibinfo {volume} {89}},\ \bibinfo
  {pages} {081407} (\bibinfo {year} {2014})}\BibitemShut {NoStop}%
\bibitem [{\citenamefont {Zhou}\ \emph {et~al.}(2013)\citenamefont {Zhou},
  \citenamefont {Jiang}, \citenamefont {Niu},\ and\ \citenamefont
  {Shi}}]{Zhou_2013}%
  \BibitemOpen
  \bibfield  {author} {\bibinfo {author} {\bibfnamefont {J.-H.}\ \bibnamefont
  {Zhou}}, \bibinfo {author} {\bibfnamefont {H.}~\bibnamefont {Jiang}},
  \bibinfo {author} {\bibfnamefont {Q.}~\bibnamefont {Niu}}, \ and\ \bibinfo
  {author} {\bibfnamefont {J.-R.}\ \bibnamefont {Shi}},\ }\href {\doibase
  10.1088/0256-307x/30/2/027101} {\bibfield  {journal} {\bibinfo  {journal}
  {Chinese Physics Letters}\ }\textbf {\bibinfo {volume} {30}},\ \bibinfo
  {pages} {027101} (\bibinfo {year} {2013})}\BibitemShut {NoStop}%
\bibitem [{\citenamefont {Burkov}(2015)}]{Burkov_2015}%
  \BibitemOpen
  \bibfield  {author} {\bibinfo {author} {\bibfnamefont {A.~A.}\ \bibnamefont
  {Burkov}},\ }\href {\doibase 10.1088/0953-8984/27/11/113201} {\bibfield
  {journal} {\bibinfo  {journal} {Journal of Physics: Condensed Matter}\
  }\textbf {\bibinfo {volume} {27}},\ \bibinfo {pages} {113201} (\bibinfo
  {year} {2015})}\BibitemShut {NoStop}%
\bibitem [{\citenamefont {Ma}\ and\ \citenamefont
  {Pesin}(2015)}]{PhysRevB.92.235205}%
  \BibitemOpen
  \bibfield  {author} {\bibinfo {author} {\bibfnamefont {J.}~\bibnamefont
  {Ma}}\ and\ \bibinfo {author} {\bibfnamefont {D.~A.}\ \bibnamefont {Pesin}},\
  }\href {\doibase 10.1103/PhysRevB.92.235205} {\bibfield  {journal} {\bibinfo
  {journal} {Phys. Rev. B}\ }\textbf {\bibinfo {volume} {92}},\ \bibinfo
  {pages} {235205} (\bibinfo {year} {2015})}\BibitemShut {NoStop}%
\bibitem [{\citenamefont {Zhong}\ \emph {et~al.}(2016)\citenamefont {Zhong},
  \citenamefont {Moore},\ and\ \citenamefont {Souza}}]{PhysRevLett.116.077201}%
  \BibitemOpen
  \bibfield  {author} {\bibinfo {author} {\bibfnamefont {S.}~\bibnamefont
  {Zhong}}, \bibinfo {author} {\bibfnamefont {J.~E.}\ \bibnamefont {Moore}}, \
  and\ \bibinfo {author} {\bibfnamefont {I.}~\bibnamefont {Souza}},\ }\href
  {\doibase 10.1103/PhysRevLett.116.077201} {\bibfield  {journal} {\bibinfo
  {journal} {Phys. Rev. Lett.}\ }\textbf {\bibinfo {volume} {116}},\ \bibinfo
  {pages} {077201} (\bibinfo {year} {2016})}\BibitemShut {NoStop}%
\bibitem [{\citenamefont {Spivak}\ and\ \citenamefont
  {Andreev}(2016)}]{PhysRevB.93.085107}%
  \BibitemOpen
  \bibfield  {author} {\bibinfo {author} {\bibfnamefont {B.~Z.}\ \bibnamefont
  {Spivak}}\ and\ \bibinfo {author} {\bibfnamefont {A.~V.}\ \bibnamefont
  {Andreev}},\ }\href {\doibase 10.1103/PhysRevB.93.085107} {\bibfield
  {journal} {\bibinfo  {journal} {Phys. Rev. B}\ }\textbf {\bibinfo {volume}
  {93}},\ \bibinfo {pages} {085107} (\bibinfo {year} {2016})}\BibitemShut
  {NoStop}%
\bibitem [{\citenamefont {Hirschberger}\ \emph {et~al.}(2016)\citenamefont
  {Hirschberger}, \citenamefont {Kushwaha}, \citenamefont {Wang}, \citenamefont
  {Gibson}, \citenamefont {Liang}, \citenamefont {Belvin}, \citenamefont
  {Bernevig}, \citenamefont {Cava},\ and\ \citenamefont
  {Ong}}]{hirschberger2016chiral}%
  \BibitemOpen
  \bibfield  {author} {\bibinfo {author} {\bibfnamefont {M.}~\bibnamefont
  {Hirschberger}}, \bibinfo {author} {\bibfnamefont {S.}~\bibnamefont
  {Kushwaha}}, \bibinfo {author} {\bibfnamefont {Z.}~\bibnamefont {Wang}},
  \bibinfo {author} {\bibfnamefont {Q.}~\bibnamefont {Gibson}}, \bibinfo
  {author} {\bibfnamefont {S.}~\bibnamefont {Liang}}, \bibinfo {author}
  {\bibfnamefont {C.~A.}\ \bibnamefont {Belvin}}, \bibinfo {author}
  {\bibfnamefont {B.~A.}\ \bibnamefont {Bernevig}}, \bibinfo {author}
  {\bibfnamefont {R.~J.}\ \bibnamefont {Cava}}, \ and\ \bibinfo {author}
  {\bibfnamefont {N.~P.}\ \bibnamefont {Ong}},\ }\href
  {https://www.nature.com/articles/nmat4684} {\bibfield  {journal} {\bibinfo
  {journal} {Nature materials}\ }\textbf {\bibinfo {volume} {15}},\ \bibinfo
  {pages} {1161} (\bibinfo {year} {2016})}\BibitemShut {NoStop}%
\bibitem [{\citenamefont {Huang}\ \emph
  {et~al.}(2015{\natexlab{b}})\citenamefont {Huang}, \citenamefont {Zhao},
  \citenamefont {Long}, \citenamefont {Wang}, \citenamefont {Chen},
  \citenamefont {Yang}, \citenamefont {Liang}, \citenamefont {Xue},
  \citenamefont {Weng}, \citenamefont {Fang}, \citenamefont {Dai},\ and\
  \citenamefont {Chen}}]{PhysRevX.5.031023}%
  \BibitemOpen
  \bibfield  {author} {\bibinfo {author} {\bibfnamefont {X.}~\bibnamefont
  {Huang}}, \bibinfo {author} {\bibfnamefont {L.}~\bibnamefont {Zhao}},
  \bibinfo {author} {\bibfnamefont {Y.}~\bibnamefont {Long}}, \bibinfo {author}
  {\bibfnamefont {P.}~\bibnamefont {Wang}}, \bibinfo {author} {\bibfnamefont
  {D.}~\bibnamefont {Chen}}, \bibinfo {author} {\bibfnamefont {Z.}~\bibnamefont
  {Yang}}, \bibinfo {author} {\bibfnamefont {H.}~\bibnamefont {Liang}},
  \bibinfo {author} {\bibfnamefont {M.}~\bibnamefont {Xue}}, \bibinfo {author}
  {\bibfnamefont {H.}~\bibnamefont {Weng}}, \bibinfo {author} {\bibfnamefont
  {Z.}~\bibnamefont {Fang}}, \bibinfo {author} {\bibfnamefont {X.}~\bibnamefont
  {Dai}}, \ and\ \bibinfo {author} {\bibfnamefont {G.}~\bibnamefont {Chen}},\
  }\href {\doibase 10.1103/PhysRevX.5.031023} {\bibfield  {journal} {\bibinfo
  {journal} {Phys. Rev. X}\ }\textbf {\bibinfo {volume} {5}},\ \bibinfo {pages}
  {031023} (\bibinfo {year} {2015}{\natexlab{b}})}\BibitemShut {NoStop}%
\bibitem [{\citenamefont {Shekhar}\ \emph {et~al.}(2015)\citenamefont
  {Shekhar}, \citenamefont {Nayak}, \citenamefont {Sun}, \citenamefont
  {Schmidt}, \citenamefont {Nicklas}, \citenamefont {Leermakers}, \citenamefont
  {Zeitler}, \citenamefont {Skourski}, \citenamefont {Wosnitza}, \citenamefont
  {Liu} \emph {et~al.}}]{shekhar2015extremely}%
  \BibitemOpen
  \bibfield  {author} {\bibinfo {author} {\bibfnamefont {C.}~\bibnamefont
  {Shekhar}}, \bibinfo {author} {\bibfnamefont {A.~K.}\ \bibnamefont {Nayak}},
  \bibinfo {author} {\bibfnamefont {Y.}~\bibnamefont {Sun}}, \bibinfo {author}
  {\bibfnamefont {M.}~\bibnamefont {Schmidt}}, \bibinfo {author} {\bibfnamefont
  {M.}~\bibnamefont {Nicklas}}, \bibinfo {author} {\bibfnamefont
  {I.}~\bibnamefont {Leermakers}}, \bibinfo {author} {\bibfnamefont
  {U.}~\bibnamefont {Zeitler}}, \bibinfo {author} {\bibfnamefont
  {Y.}~\bibnamefont {Skourski}}, \bibinfo {author} {\bibfnamefont
  {J.}~\bibnamefont {Wosnitza}}, \bibinfo {author} {\bibfnamefont
  {Z.}~\bibnamefont {Liu}},  \emph {et~al.},\ }\href
  {https://www.nature.com/articles/nphys3372} {\bibfield  {journal} {\bibinfo
  {journal} {Nature Physics}\ }\textbf {\bibinfo {volume} {11}},\ \bibinfo
  {pages} {645} (\bibinfo {year} {2015})}\BibitemShut {NoStop}%
\bibitem [{\citenamefont {Du}\ \emph {et~al.}(2016)\citenamefont {Du},
  \citenamefont {Wang}, \citenamefont {Chen}, \citenamefont {Mao},
  \citenamefont {Khan}, \citenamefont {Xu}, \citenamefont {Zhou}, \citenamefont
  {Zhang}, \citenamefont {Yang}, \citenamefont {Chen} \emph
  {et~al.}}]{du2016large}%
  \BibitemOpen
  \bibfield  {author} {\bibinfo {author} {\bibfnamefont {J.}~\bibnamefont
  {Du}}, \bibinfo {author} {\bibfnamefont {H.}~\bibnamefont {Wang}}, \bibinfo
  {author} {\bibfnamefont {Q.}~\bibnamefont {Chen}}, \bibinfo {author}
  {\bibfnamefont {Q.}~\bibnamefont {Mao}}, \bibinfo {author} {\bibfnamefont
  {R.}~\bibnamefont {Khan}}, \bibinfo {author} {\bibfnamefont {B.}~\bibnamefont
  {Xu}}, \bibinfo {author} {\bibfnamefont {Y.}~\bibnamefont {Zhou}}, \bibinfo
  {author} {\bibfnamefont {Y.}~\bibnamefont {Zhang}}, \bibinfo {author}
  {\bibfnamefont {J.}~\bibnamefont {Yang}}, \bibinfo {author} {\bibfnamefont
  {B.}~\bibnamefont {Chen}},  \emph {et~al.},\ }\href
  {https://link.springer.com/article/10.1007/s11433-016-5798-4} {\bibfield
  {journal} {\bibinfo  {journal} {Science China Physics, Mechanics \&
  Astronomy}\ }\textbf {\bibinfo {volume} {59}},\ \bibinfo {pages} {657406}
  (\bibinfo {year} {2016})}\BibitemShut {NoStop}%
\bibitem [{\citenamefont {Wang}\ \emph
  {et~al.}(2016{\natexlab{a}})\citenamefont {Wang}, \citenamefont {Zheng},
  \citenamefont {Shen}, \citenamefont {Lu}, \citenamefont {Fang}, \citenamefont
  {Sheng}, \citenamefont {Zhou}, \citenamefont {Yang}, \citenamefont {Li},
  \citenamefont {Feng},\ and\ \citenamefont {Xu}}]{PhysRevB.93.121112}%
  \BibitemOpen
  \bibfield  {author} {\bibinfo {author} {\bibfnamefont {Z.}~\bibnamefont
  {Wang}}, \bibinfo {author} {\bibfnamefont {Y.}~\bibnamefont {Zheng}},
  \bibinfo {author} {\bibfnamefont {Z.}~\bibnamefont {Shen}}, \bibinfo {author}
  {\bibfnamefont {Y.}~\bibnamefont {Lu}}, \bibinfo {author} {\bibfnamefont
  {H.}~\bibnamefont {Fang}}, \bibinfo {author} {\bibfnamefont {F.}~\bibnamefont
  {Sheng}}, \bibinfo {author} {\bibfnamefont {Y.}~\bibnamefont {Zhou}},
  \bibinfo {author} {\bibfnamefont {X.}~\bibnamefont {Yang}}, \bibinfo {author}
  {\bibfnamefont {Y.}~\bibnamefont {Li}}, \bibinfo {author} {\bibfnamefont
  {C.}~\bibnamefont {Feng}}, \ and\ \bibinfo {author} {\bibfnamefont {Z.-A.}\
  \bibnamefont {Xu}},\ }\href {\doibase 10.1103/PhysRevB.93.121112} {\bibfield
  {journal} {\bibinfo  {journal} {Phys. Rev. B}\ }\textbf {\bibinfo {volume}
  {93}},\ \bibinfo {pages} {121112} (\bibinfo {year}
  {2016}{\natexlab{a}})}\BibitemShut {NoStop}%
\bibitem [{\citenamefont {Zhang}\ \emph {et~al.}(2016)\citenamefont {Zhang},
  \citenamefont {Xu}, \citenamefont {Belopolski}, \citenamefont {Yuan},
  \citenamefont {Lin}, \citenamefont {Tong}, \citenamefont {Bian},
  \citenamefont {Alidoust}, \citenamefont {Lee}, \citenamefont {Huang} \emph
  {et~al.}}]{zhang2016signatures}%
  \BibitemOpen
  \bibfield  {author} {\bibinfo {author} {\bibfnamefont {C.-L.}\ \bibnamefont
  {Zhang}}, \bibinfo {author} {\bibfnamefont {S.-Y.}\ \bibnamefont {Xu}},
  \bibinfo {author} {\bibfnamefont {I.}~\bibnamefont {Belopolski}}, \bibinfo
  {author} {\bibfnamefont {Z.}~\bibnamefont {Yuan}}, \bibinfo {author}
  {\bibfnamefont {Z.}~\bibnamefont {Lin}}, \bibinfo {author} {\bibfnamefont
  {B.}~\bibnamefont {Tong}}, \bibinfo {author} {\bibfnamefont {G.}~\bibnamefont
  {Bian}}, \bibinfo {author} {\bibfnamefont {N.}~\bibnamefont {Alidoust}},
  \bibinfo {author} {\bibfnamefont {C.-C.}\ \bibnamefont {Lee}}, \bibinfo
  {author} {\bibfnamefont {S.-M.}\ \bibnamefont {Huang}},  \emph {et~al.},\
  }\href {https://www.nature.com/articles/ncomms10735} {\bibfield  {journal}
  {\bibinfo  {journal} {Nature communications}\ }\textbf {\bibinfo {volume}
  {7}},\ \bibinfo {pages} {1} (\bibinfo {year} {2016})}\BibitemShut {NoStop}%
\bibitem [{\citenamefont {Nielsen}\ and\ \citenamefont
  {Ninomiya}(1981{\natexlab{a}})}]{NIELSEN198120}%
  \BibitemOpen
  \bibfield  {author} {\bibinfo {author} {\bibfnamefont {H.}~\bibnamefont
  {Nielsen}}\ and\ \bibinfo {author} {\bibfnamefont {M.}~\bibnamefont
  {Ninomiya}},\ }\href {\doibase https://doi.org/10.1016/0550-3213(81)90361-8}
  {\bibfield  {journal} {\bibinfo  {journal} {Nuclear Physics B}\ }\textbf
  {\bibinfo {volume} {185}},\ \bibinfo {pages} {20} (\bibinfo {year}
  {1981}{\natexlab{a}})}\BibitemShut {NoStop}%
\bibitem [{\citenamefont {Nielsen}\ and\ \citenamefont
  {Ninomiya}(1981{\natexlab{b}})}]{NIELSEN1981173}%
  \BibitemOpen
  \bibfield  {author} {\bibinfo {author} {\bibfnamefont {H.}~\bibnamefont
  {Nielsen}}\ and\ \bibinfo {author} {\bibfnamefont {M.}~\bibnamefont
  {Ninomiya}},\ }\href {\doibase https://doi.org/10.1016/0550-3213(81)90524-1}
  {\bibfield  {journal} {\bibinfo  {journal} {Nuclear Physics B}\ }\textbf
  {\bibinfo {volume} {193}},\ \bibinfo {pages} {173} (\bibinfo {year}
  {1981}{\natexlab{b}})}\BibitemShut {NoStop}%
\bibitem [{\citenamefont {Nielsen}\ and\ \citenamefont
  {Ninomiya}(1983)}]{NIELSEN1983389}%
  \BibitemOpen
  \bibfield  {author} {\bibinfo {author} {\bibfnamefont {H.}~\bibnamefont
  {Nielsen}}\ and\ \bibinfo {author} {\bibfnamefont {M.}~\bibnamefont
  {Ninomiya}},\ }\href {\doibase https://doi.org/10.1016/0370-2693(83)91529-0}
  {\bibfield  {journal} {\bibinfo  {journal} {Physics Letters B}\ }\textbf
  {\bibinfo {volume} {130}},\ \bibinfo {pages} {389} (\bibinfo {year}
  {1983})}\BibitemShut {NoStop}%
\bibitem [{\citenamefont {Morali}\ \emph {et~al.}(2019)\citenamefont {Morali},
  \citenamefont {Batabyal}, \citenamefont {Nag}, \citenamefont {Liu},
  \citenamefont {Xu}, \citenamefont {Sun}, \citenamefont {Yan}, \citenamefont
  {Felser}, \citenamefont {Avraham},\ and\ \citenamefont
  {Beidenkopf}}]{Morali1286}%
  \BibitemOpen
  \bibfield  {author} {\bibinfo {author} {\bibfnamefont {N.}~\bibnamefont
  {Morali}}, \bibinfo {author} {\bibfnamefont {R.}~\bibnamefont {Batabyal}},
  \bibinfo {author} {\bibfnamefont {P.~K.}\ \bibnamefont {Nag}}, \bibinfo
  {author} {\bibfnamefont {E.}~\bibnamefont {Liu}}, \bibinfo {author}
  {\bibfnamefont {Q.}~\bibnamefont {Xu}}, \bibinfo {author} {\bibfnamefont
  {Y.}~\bibnamefont {Sun}}, \bibinfo {author} {\bibfnamefont {B.}~\bibnamefont
  {Yan}}, \bibinfo {author} {\bibfnamefont {C.}~\bibnamefont {Felser}},
  \bibinfo {author} {\bibfnamefont {N.}~\bibnamefont {Avraham}}, \ and\
  \bibinfo {author} {\bibfnamefont {H.}~\bibnamefont {Beidenkopf}},\ }\href
  {\doibase 10.1126/science.aav2334} {\bibfield  {journal} {\bibinfo  {journal}
  {Science}\ }\textbf {\bibinfo {volume} {365}},\ \bibinfo {pages} {1286}
  (\bibinfo {year} {2019})}\BibitemShut {NoStop}%
\bibitem [{\citenamefont {Yang}\ \emph {et~al.}(2019)\citenamefont {Yang},
  \citenamefont {Yang}, \citenamefont {Liu}, \citenamefont {Sun}, \citenamefont
  {Chen}, \citenamefont {Peng}, \citenamefont {Schmidt}, \citenamefont
  {Prabhakaran}, \citenamefont {Bernevig}, \citenamefont {Felser} \emph
  {et~al.}}]{yang2019topological}%
  \BibitemOpen
  \bibfield  {author} {\bibinfo {author} {\bibfnamefont {H.}~\bibnamefont
  {Yang}}, \bibinfo {author} {\bibfnamefont {L.}~\bibnamefont {Yang}}, \bibinfo
  {author} {\bibfnamefont {Z.}~\bibnamefont {Liu}}, \bibinfo {author}
  {\bibfnamefont {Y.}~\bibnamefont {Sun}}, \bibinfo {author} {\bibfnamefont
  {C.}~\bibnamefont {Chen}}, \bibinfo {author} {\bibfnamefont {H.}~\bibnamefont
  {Peng}}, \bibinfo {author} {\bibfnamefont {M.}~\bibnamefont {Schmidt}},
  \bibinfo {author} {\bibfnamefont {D.}~\bibnamefont {Prabhakaran}}, \bibinfo
  {author} {\bibfnamefont {B.~A.}\ \bibnamefont {Bernevig}}, \bibinfo {author}
  {\bibfnamefont {C.}~\bibnamefont {Felser}},  \emph {et~al.},\ }\href
  {https://www.nature.com/articles/s41467-019-11491-4} {\bibfield  {journal}
  {\bibinfo  {journal} {Nature communications}\ }\textbf {\bibinfo {volume}
  {10}},\ \bibinfo {pages} {1} (\bibinfo {year} {2019})}\BibitemShut {NoStop}%
\bibitem [{\citenamefont {Ekahana}\ \emph {et~al.}(2020)\citenamefont
  {Ekahana}, \citenamefont {Li}, \citenamefont {Sun}, \citenamefont {Namiki},
  \citenamefont {Yang}, \citenamefont {Jiang}, \citenamefont {Yang},
  \citenamefont {Shi}, \citenamefont {Zhang}, \citenamefont {Pei},
  \citenamefont {Chen}, \citenamefont {Sasagawa}, \citenamefont {Felser},
  \citenamefont {Yan}, \citenamefont {Liu},\ and\ \citenamefont
  {Chen}}]{PhysRevB.102.085126}%
  \BibitemOpen
  \bibfield  {author} {\bibinfo {author} {\bibfnamefont {S.~A.}\ \bibnamefont
  {Ekahana}}, \bibinfo {author} {\bibfnamefont {Y.~W.}\ \bibnamefont {Li}},
  \bibinfo {author} {\bibfnamefont {Y.}~\bibnamefont {Sun}}, \bibinfo {author}
  {\bibfnamefont {H.}~\bibnamefont {Namiki}}, \bibinfo {author} {\bibfnamefont
  {H.~F.}\ \bibnamefont {Yang}}, \bibinfo {author} {\bibfnamefont
  {J.}~\bibnamefont {Jiang}}, \bibinfo {author} {\bibfnamefont {L.~X.}\
  \bibnamefont {Yang}}, \bibinfo {author} {\bibfnamefont {W.~J.}\ \bibnamefont
  {Shi}}, \bibinfo {author} {\bibfnamefont {C.~F.}\ \bibnamefont {Zhang}},
  \bibinfo {author} {\bibfnamefont {D.}~\bibnamefont {Pei}}, \bibinfo {author}
  {\bibfnamefont {C.}~\bibnamefont {Chen}}, \bibinfo {author} {\bibfnamefont
  {T.}~\bibnamefont {Sasagawa}}, \bibinfo {author} {\bibfnamefont
  {C.}~\bibnamefont {Felser}}, \bibinfo {author} {\bibfnamefont {B.~H.}\
  \bibnamefont {Yan}}, \bibinfo {author} {\bibfnamefont {Z.~K.}\ \bibnamefont
  {Liu}}, \ and\ \bibinfo {author} {\bibfnamefont {Y.~L.}\ \bibnamefont
  {Chen}},\ }\href {\doibase 10.1103/PhysRevB.102.085126} {\bibfield  {journal}
  {\bibinfo  {journal} {Phys. Rev. B}\ }\textbf {\bibinfo {volume} {102}},\
  \bibinfo {pages} {085126} (\bibinfo {year} {2020})}\BibitemShut {NoStop}%
\bibitem [{\citenamefont {Chen}\ \emph
  {et~al.}(2018{\natexlab{a}})\citenamefont {Chen}, \citenamefont {Luo},
  \citenamefont {Li},\ and\ \citenamefont
  {Zilberberg}}]{PhysRevLett.121.166802}%
  \BibitemOpen
  \bibfield  {author} {\bibinfo {author} {\bibfnamefont {W.}~\bibnamefont
  {Chen}}, \bibinfo {author} {\bibfnamefont {K.}~\bibnamefont {Luo}}, \bibinfo
  {author} {\bibfnamefont {L.}~\bibnamefont {Li}}, \ and\ \bibinfo {author}
  {\bibfnamefont {O.}~\bibnamefont {Zilberberg}},\ }\href {\doibase
  10.1103/PhysRevLett.121.166802} {\bibfield  {journal} {\bibinfo  {journal}
  {Phys. Rev. Lett.}\ }\textbf {\bibinfo {volume} {121}},\ \bibinfo {pages}
  {166802} (\bibinfo {year} {2018}{\natexlab{a}})}\BibitemShut {NoStop}%
\bibitem [{\citenamefont {Chen}\ \emph {et~al.}(2020)\citenamefont {Chen},
  \citenamefont {Zilberberg},\ and\ \citenamefont
  {Chen}}]{PhysRevB.101.125407}%
  \BibitemOpen
  \bibfield  {author} {\bibinfo {author} {\bibfnamefont {G.}~\bibnamefont
  {Chen}}, \bibinfo {author} {\bibfnamefont {O.}~\bibnamefont {Zilberberg}}, \
  and\ \bibinfo {author} {\bibfnamefont {W.}~\bibnamefont {Chen}},\ }\href
  {\doibase 10.1103/PhysRevB.101.125407} {\bibfield  {journal} {\bibinfo
  {journal} {Phys. Rev. B}\ }\textbf {\bibinfo {volume} {101}},\ \bibinfo
  {pages} {125407} (\bibinfo {year} {2020})}\BibitemShut {NoStop}%
\bibitem [{\citenamefont {Zheng}\ \emph {et~al.}(2020)\citenamefont {Zheng},
  \citenamefont {Chen},\ and\ \citenamefont {Xing}}]{zheng2020andreev}%
  \BibitemOpen
  \bibfield  {author} {\bibinfo {author} {\bibfnamefont {Y.}~\bibnamefont
  {Zheng}}, \bibinfo {author} {\bibfnamefont {W.}~\bibnamefont {Chen}}, \ and\
  \bibinfo {author} {\bibfnamefont {D.}~\bibnamefont {Xing}},\ }\href@noop {}
  {\bibfield  {journal} {\bibinfo  {journal} {arXiv preprint arXiv:2012.08066}\
  } (\bibinfo {year} {2020})}\BibitemShut {NoStop}%
\bibitem [{\citenamefont {Young}\ and\ \citenamefont
  {Kim}(2008)}]{2008Quantum}%
  \BibitemOpen
  \bibfield  {author} {\bibinfo {author} {\bibfnamefont {A.~F.}\ \bibnamefont
  {Young}}\ and\ \bibinfo {author} {\bibfnamefont {P.}~\bibnamefont {Kim}},\
  }\href {\doibase 10.1038/nphys1198} {\bibfield  {journal} {\bibinfo
  {journal} {Nature Physics}\ }\textbf {\bibinfo {volume} {5}},\ \bibinfo
  {pages} {222} (\bibinfo {year} {2008})}\BibitemShut {NoStop}%
\bibitem [{\citenamefont {Rickhaus}\ \emph {et~al.}(2013)\citenamefont
  {Rickhaus}, \citenamefont {Maurand}, \citenamefont {Liu}, \citenamefont
  {Weiss}, \citenamefont {Richter},\ and\ \citenamefont
  {Sch{\"o}Nenberger}}]{2013Ballistic}%
  \BibitemOpen
  \bibfield  {author} {\bibinfo {author} {\bibfnamefont {P.}~\bibnamefont
  {Rickhaus}}, \bibinfo {author} {\bibfnamefont {R.}~\bibnamefont {Maurand}},
  \bibinfo {author} {\bibfnamefont {M.~H.}\ \bibnamefont {Liu}}, \bibinfo
  {author} {\bibfnamefont {M.}~\bibnamefont {Weiss}}, \bibinfo {author}
  {\bibfnamefont {K.}~\bibnamefont {Richter}}, \ and\ \bibinfo {author}
  {\bibfnamefont {C.}~\bibnamefont {Sch{\"o}Nenberger}},\ }\href {\doibase
  10.1038/ncomms3342} {\bibfield  {journal} {\bibinfo  {journal} {Nature
  Communications}\ }\textbf {\bibinfo {volume} {4}},\ \bibinfo {pages} {2342}
  (\bibinfo {year} {2013})}\BibitemShut {NoStop}%
\bibitem [{\citenamefont {Grushina}\ \emph {et~al.}(2013)\citenamefont
  {Grushina}, \citenamefont {Ki},\ and\ \citenamefont {Morpurgo}}]{2013A}%
  \BibitemOpen
  \bibfield  {author} {\bibinfo {author} {\bibfnamefont {A.~L.}\ \bibnamefont
  {Grushina}}, \bibinfo {author} {\bibfnamefont {D.-K.}\ \bibnamefont {Ki}}, \
  and\ \bibinfo {author} {\bibfnamefont {A.~F.}\ \bibnamefont {Morpurgo}},\
  }\href {\doibase 10.1063/1.4807888} {\bibfield  {journal} {\bibinfo
  {journal} {Appl. Phys. Lett.}\ }\textbf {\bibinfo {volume} {102}},\ \bibinfo
  {pages} {223102} (\bibinfo {year} {2013})}\BibitemShut {NoStop}%
\bibitem [{\citenamefont {Oksanen}\ \emph {et~al.}(2014)\citenamefont
  {Oksanen}, \citenamefont {Uppstu}, \citenamefont {Laitinen}, \citenamefont
  {Cox}, \citenamefont {Craciun}, \citenamefont {Russo}, \citenamefont
  {Harju},\ and\ \citenamefont {Hakonen}}]{Oksanen2014Single}%
  \BibitemOpen
  \bibfield  {author} {\bibinfo {author} {\bibfnamefont {M.}~\bibnamefont
  {Oksanen}}, \bibinfo {author} {\bibfnamefont {A.}~\bibnamefont {Uppstu}},
  \bibinfo {author} {\bibfnamefont {A.}~\bibnamefont {Laitinen}}, \bibinfo
  {author} {\bibfnamefont {D.~J.}\ \bibnamefont {Cox}}, \bibinfo {author}
  {\bibfnamefont {M.~F.}\ \bibnamefont {Craciun}}, \bibinfo {author}
  {\bibfnamefont {S.}~\bibnamefont {Russo}}, \bibinfo {author} {\bibfnamefont
  {A.}~\bibnamefont {Harju}}, \ and\ \bibinfo {author} {\bibfnamefont
  {P.}~\bibnamefont {Hakonen}},\ }\href {\doibase 10.1103/PhysRevB.89.121414}
  {\bibfield  {journal} {\bibinfo  {journal} {Phys. Rev. B}\ }\textbf {\bibinfo
  {volume} {89}},\ \bibinfo {pages} {121414(R)} (\bibinfo {year}
  {2014})}\BibitemShut {NoStop}%
\bibitem [{\citenamefont {Campos}\ \emph {et~al.}(2012)\citenamefont {Campos},
  \citenamefont {Young}, \citenamefont {Surakitbovorn}, \citenamefont
  {Watanabe}, \citenamefont {Taniguchi},\ and\ \citenamefont
  {Jarillo-Herrero}}]{0Quantum}%
  \BibitemOpen
  \bibfield  {author} {\bibinfo {author} {\bibfnamefont {L.~C.}\ \bibnamefont
  {Campos}}, \bibinfo {author} {\bibfnamefont {A.~F.}\ \bibnamefont {Young}},
  \bibinfo {author} {\bibfnamefont {K.}~\bibnamefont {Surakitbovorn}}, \bibinfo
  {author} {\bibfnamefont {K.}~\bibnamefont {Watanabe}}, \bibinfo {author}
  {\bibfnamefont {T.}~\bibnamefont {Taniguchi}}, \ and\ \bibinfo {author}
  {\bibfnamefont {P.}~\bibnamefont {Jarillo-Herrero}},\ }\href {\doibase
  10.1038/ncomms2243} {\bibfield  {journal} {\bibinfo  {journal} {Nature
  Communications}\ }\textbf {\bibinfo {volume} {3}},\ \bibinfo {pages} {1239}
  (\bibinfo {year} {2012})}\BibitemShut {NoStop}%
\bibitem [{\citenamefont {Varlet}\ \emph {et~al.}(2014)\citenamefont {Varlet},
  \citenamefont {Liu}, \citenamefont {Krueckl}, \citenamefont {Bischoff},
  \citenamefont {Simonet}, \citenamefont {Watanabe}, \citenamefont {Taniguchi},
  \citenamefont {Richter}, \citenamefont {Ensslin},\ and\ \citenamefont
  {Ihn}}]{PhysRevLett.113.116601}%
  \BibitemOpen
  \bibfield  {author} {\bibinfo {author} {\bibfnamefont {A.}~\bibnamefont
  {Varlet}}, \bibinfo {author} {\bibfnamefont {M.-H.}\ \bibnamefont {Liu}},
  \bibinfo {author} {\bibfnamefont {V.}~\bibnamefont {Krueckl}}, \bibinfo
  {author} {\bibfnamefont {D.}~\bibnamefont {Bischoff}}, \bibinfo {author}
  {\bibfnamefont {P.}~\bibnamefont {Simonet}}, \bibinfo {author} {\bibfnamefont
  {K.}~\bibnamefont {Watanabe}}, \bibinfo {author} {\bibfnamefont
  {T.}~\bibnamefont {Taniguchi}}, \bibinfo {author} {\bibfnamefont
  {K.}~\bibnamefont {Richter}}, \bibinfo {author} {\bibfnamefont
  {K.}~\bibnamefont {Ensslin}}, \ and\ \bibinfo {author} {\bibfnamefont
  {T.}~\bibnamefont {Ihn}},\ }\href {\doibase 10.1103/PhysRevLett.113.116601}
  {\bibfield  {journal} {\bibinfo  {journal} {Phys. Rev. Lett.}\ }\textbf
  {\bibinfo {volume} {113}},\ \bibinfo {pages} {116601} (\bibinfo {year}
  {2014})}\BibitemShut {NoStop}%
\bibitem [{\citenamefont {Karalic}\ \emph {et~al.}(2020)\citenamefont
  {Karalic}, \citenamefont {Trkalj}, \citenamefont {Masseroni}, \citenamefont
  {Chen},\ and\ \citenamefont {Zilberberg}}]{PhysRevX.10.031007}%
  \BibitemOpen
  \bibfield  {author} {\bibinfo {author} {\bibfnamefont {M.}~\bibnamefont
  {Karalic}}, \bibinfo {author} {\bibfnamefont {A.}~\bibnamefont {Trkalj}},
  \bibinfo {author} {\bibfnamefont {M.}~\bibnamefont {Masseroni}}, \bibinfo
  {author} {\bibfnamefont {W.}~\bibnamefont {Chen}}, \ and\ \bibinfo {author}
  {\bibfnamefont {O.}~\bibnamefont {Zilberberg}},\ }\href {\doibase
  10.1103/PhysRevX.10.031007} {\bibfield  {journal} {\bibinfo  {journal} {Phys.
  Rev. X}\ }\textbf {\bibinfo {volume} {10}},\ \bibinfo {pages} {031007}
  (\bibinfo {year} {2020})}\BibitemShut {NoStop}%
\bibitem [{\citenamefont {Chen}\ \emph {et~al.}(2013)\citenamefont {Chen},
  \citenamefont {Jiang}, \citenamefont {Shen}, \citenamefont {Sheng},
  \citenamefont {Wang},\ and\ \citenamefont {Xing}}]{Chen_2013}%
  \BibitemOpen
  \bibfield  {author} {\bibinfo {author} {\bibfnamefont {W.}~\bibnamefont
  {Chen}}, \bibinfo {author} {\bibfnamefont {L.}~\bibnamefont {Jiang}},
  \bibinfo {author} {\bibfnamefont {R.}~\bibnamefont {Shen}}, \bibinfo {author}
  {\bibfnamefont {L.}~\bibnamefont {Sheng}}, \bibinfo {author} {\bibfnamefont
  {B.~G.}\ \bibnamefont {Wang}}, \ and\ \bibinfo {author} {\bibfnamefont
  {D.~Y.}\ \bibnamefont {Xing}},\ }\href {\doibase 10.1209/0295-5075/103/27006}
  {\bibfield  {journal} {\bibinfo  {journal} {{EPL} (Europhysics Letters)}\
  }\textbf {\bibinfo {volume} {103}},\ \bibinfo {pages} {27006} (\bibinfo
  {year} {2013})}\BibitemShut {NoStop}%
\bibitem [{\citenamefont {Datta}(1997)}]{datta1997electronic}%
  \BibitemOpen
  \bibfield  {author} {\bibinfo {author} {\bibfnamefont {S.}~\bibnamefont
  {Datta}},\ }\href@noop {} {\emph {\bibinfo {title} {Electronic transport in
  mesoscopic systems}}}\ (\bibinfo  {publisher} {Cambridge university press},\
  \bibinfo {year} {1997})\BibitemShut {NoStop}%
\bibitem [{\citenamefont {Koepernik}\ \emph {et~al.}(2016)\citenamefont
  {Koepernik}, \citenamefont {Kasinathan}, \citenamefont {Efremov},
  \citenamefont {Khim}, \citenamefont {Borisenko}, \citenamefont {B\"uchner},\
  and\ \citenamefont {van~den Brink}}]{PhysRevB.93.201101}%
  \BibitemOpen
  \bibfield  {author} {\bibinfo {author} {\bibfnamefont {K.}~\bibnamefont
  {Koepernik}}, \bibinfo {author} {\bibfnamefont {D.}~\bibnamefont
  {Kasinathan}}, \bibinfo {author} {\bibfnamefont {D.~V.}\ \bibnamefont
  {Efremov}}, \bibinfo {author} {\bibfnamefont {S.}~\bibnamefont {Khim}},
  \bibinfo {author} {\bibfnamefont {S.}~\bibnamefont {Borisenko}}, \bibinfo
  {author} {\bibfnamefont {B.}~\bibnamefont {B\"uchner}}, \ and\ \bibinfo
  {author} {\bibfnamefont {J.}~\bibnamefont {van~den Brink}},\ }\href {\doibase
  10.1103/PhysRevB.93.201101} {\bibfield  {journal} {\bibinfo  {journal} {Phys.
  Rev. B}\ }\textbf {\bibinfo {volume} {93}},\ \bibinfo {pages} {201101}
  (\bibinfo {year} {2016})}\BibitemShut {NoStop}%
\bibitem [{\citenamefont {Belopolski}\ \emph {et~al.}(2017)\citenamefont
  {Belopolski}, \citenamefont {Yu}, \citenamefont {Sanchez}, \citenamefont
  {Ishida}, \citenamefont {Chang}, \citenamefont {Zhang}, \citenamefont {Xu},
  \citenamefont {Zheng}, \citenamefont {Chang}, \citenamefont {Bian} \emph
  {et~al.}}]{belopolski2017signatures}%
  \BibitemOpen
  \bibfield  {author} {\bibinfo {author} {\bibfnamefont {I.}~\bibnamefont
  {Belopolski}}, \bibinfo {author} {\bibfnamefont {P.}~\bibnamefont {Yu}},
  \bibinfo {author} {\bibfnamefont {D.~S.}\ \bibnamefont {Sanchez}}, \bibinfo
  {author} {\bibfnamefont {Y.}~\bibnamefont {Ishida}}, \bibinfo {author}
  {\bibfnamefont {T.-R.}\ \bibnamefont {Chang}}, \bibinfo {author}
  {\bibfnamefont {S.~S.}\ \bibnamefont {Zhang}}, \bibinfo {author}
  {\bibfnamefont {S.-Y.}\ \bibnamefont {Xu}}, \bibinfo {author} {\bibfnamefont
  {H.}~\bibnamefont {Zheng}}, \bibinfo {author} {\bibfnamefont
  {G.}~\bibnamefont {Chang}}, \bibinfo {author} {\bibfnamefont
  {G.}~\bibnamefont {Bian}},  \emph {et~al.},\ }\href
  {https://www.nature.com/articles/s41467-017-00938-1} {\bibfield  {journal}
  {\bibinfo  {journal} {Nature communications}\ }\textbf {\bibinfo {volume}
  {8}},\ \bibinfo {pages} {1} (\bibinfo {year} {2017})}\BibitemShut {NoStop}%
\bibitem [{\citenamefont {Haubold}\ \emph {et~al.}(2017)\citenamefont
  {Haubold}, \citenamefont {Koepernik}, \citenamefont {Efremov}, \citenamefont
  {Khim}, \citenamefont {Fedorov}, \citenamefont {Kushnirenko}, \citenamefont
  {van~den Brink}, \citenamefont {Wurmehl}, \citenamefont {B\"uchner},
  \citenamefont {Kim}, \citenamefont {Hoesch}, \citenamefont {Sumida},
  \citenamefont {Taguchi}, \citenamefont {Yoshikawa}, \citenamefont {Kimura},
  \citenamefont {Okuda},\ and\ \citenamefont {Borisenko}}]{PhysRevB.95.241108}%
  \BibitemOpen
  \bibfield  {author} {\bibinfo {author} {\bibfnamefont {E.}~\bibnamefont
  {Haubold}}, \bibinfo {author} {\bibfnamefont {K.}~\bibnamefont {Koepernik}},
  \bibinfo {author} {\bibfnamefont {D.}~\bibnamefont {Efremov}}, \bibinfo
  {author} {\bibfnamefont {S.}~\bibnamefont {Khim}}, \bibinfo {author}
  {\bibfnamefont {A.}~\bibnamefont {Fedorov}}, \bibinfo {author} {\bibfnamefont
  {Y.}~\bibnamefont {Kushnirenko}}, \bibinfo {author} {\bibfnamefont
  {J.}~\bibnamefont {van~den Brink}}, \bibinfo {author} {\bibfnamefont
  {S.}~\bibnamefont {Wurmehl}}, \bibinfo {author} {\bibfnamefont
  {B.}~\bibnamefont {B\"uchner}}, \bibinfo {author} {\bibfnamefont {T.~K.}\
  \bibnamefont {Kim}}, \bibinfo {author} {\bibfnamefont {M.}~\bibnamefont
  {Hoesch}}, \bibinfo {author} {\bibfnamefont {K.}~\bibnamefont {Sumida}},
  \bibinfo {author} {\bibfnamefont {K.}~\bibnamefont {Taguchi}}, \bibinfo
  {author} {\bibfnamefont {T.}~\bibnamefont {Yoshikawa}}, \bibinfo {author}
  {\bibfnamefont {A.}~\bibnamefont {Kimura}}, \bibinfo {author} {\bibfnamefont
  {T.}~\bibnamefont {Okuda}}, \ and\ \bibinfo {author} {\bibfnamefont {S.~V.}\
  \bibnamefont {Borisenko}},\ }\href {\doibase 10.1103/PhysRevB.95.241108}
  {\bibfield  {journal} {\bibinfo  {journal} {Phys. Rev. B}\ }\textbf {\bibinfo
  {volume} {95}},\ \bibinfo {pages} {241108} (\bibinfo {year}
  {2017})}\BibitemShut {NoStop}%
\bibitem [{\citenamefont {Groth}\ \emph {et~al.}(2014)\citenamefont {Groth},
  \citenamefont {Wimmer}, \citenamefont {Akhmerov},\ and\ \citenamefont
  {Waintal}}]{Groth_2014}%
  \BibitemOpen
  \bibfield  {author} {\bibinfo {author} {\bibfnamefont {C.~W.}\ \bibnamefont
  {Groth}}, \bibinfo {author} {\bibfnamefont {M.}~\bibnamefont {Wimmer}},
  \bibinfo {author} {\bibfnamefont {A.~R.}\ \bibnamefont {Akhmerov}}, \ and\
  \bibinfo {author} {\bibfnamefont {X.}~\bibnamefont {Waintal}},\ }\href
  {\doibase 10.1088/1367-2630/16/6/063065} {\bibfield  {journal} {\bibinfo
  {journal} {New Journal of Physics}\ }\textbf {\bibinfo {volume} {16}},\
  \bibinfo {pages} {063065} (\bibinfo {year} {2014})}\BibitemShut {NoStop}%
\bibitem [{\citenamefont {Potter}\ \emph {et~al.}(2014)\citenamefont {Potter},
  \citenamefont {Kimchi},\ and\ \citenamefont
  {Vishwanath}}]{potter2014quantum}%
  \BibitemOpen
  \bibfield  {author} {\bibinfo {author} {\bibfnamefont {A.~C.}\ \bibnamefont
  {Potter}}, \bibinfo {author} {\bibfnamefont {I.}~\bibnamefont {Kimchi}}, \
  and\ \bibinfo {author} {\bibfnamefont {A.}~\bibnamefont {Vishwanath}},\
  }\href {https://www.nature.com/articles/ncomms6161} {\bibfield  {journal}
  {\bibinfo  {journal} {Nature communications}\ }\textbf {\bibinfo {volume}
  {5}},\ \bibinfo {pages} {1} (\bibinfo {year} {2014})}\BibitemShut {NoStop}%
\bibitem [{\citenamefont {Li}\ \emph {et~al.}(2020)\citenamefont {Li},
  \citenamefont {Wang}, \citenamefont {Li}, \citenamefont {Zheng},
  \citenamefont {Brinkman}, \citenamefont {Yu},\ and\ \citenamefont
  {Liao}}]{li2020fermi}%
  \BibitemOpen
  \bibfield  {author} {\bibinfo {author} {\bibfnamefont {C.-Z.}\ \bibnamefont
  {Li}}, \bibinfo {author} {\bibfnamefont {A.-Q.}\ \bibnamefont {Wang}},
  \bibinfo {author} {\bibfnamefont {C.}~\bibnamefont {Li}}, \bibinfo {author}
  {\bibfnamefont {W.-Z.}\ \bibnamefont {Zheng}}, \bibinfo {author}
  {\bibfnamefont {A.}~\bibnamefont {Brinkman}}, \bibinfo {author}
  {\bibfnamefont {D.-P.}\ \bibnamefont {Yu}}, \ and\ \bibinfo {author}
  {\bibfnamefont {Z.-M.}\ \bibnamefont {Liao}},\ }\href
  {https://www.nature.com/articles/s41467-020-15010-8} {\bibfield  {journal}
  {\bibinfo  {journal} {Nature communications}\ }\textbf {\bibinfo {volume}
  {11}},\ \bibinfo {pages} {1} (\bibinfo {year} {2020})}\BibitemShut {NoStop}%
\bibitem [{\citenamefont {Chen}\ \emph
  {et~al.}(2018{\natexlab{b}})\citenamefont {Chen}, \citenamefont {Park},
  \citenamefont {Gill}, \citenamefont {Xiao}, \citenamefont {Reig-i Plessis},
  \citenamefont {MacDougall}, \citenamefont {Gilbert},\ and\ \citenamefont
  {Mason}}]{chen2018finite}%
  \BibitemOpen
  \bibfield  {author} {\bibinfo {author} {\bibfnamefont {A.~Q.}\ \bibnamefont
  {Chen}}, \bibinfo {author} {\bibfnamefont {M.~J.}\ \bibnamefont {Park}},
  \bibinfo {author} {\bibfnamefont {S.~T.}\ \bibnamefont {Gill}}, \bibinfo
  {author} {\bibfnamefont {Y.}~\bibnamefont {Xiao}}, \bibinfo {author}
  {\bibfnamefont {D.}~\bibnamefont {Reig-i Plessis}}, \bibinfo {author}
  {\bibfnamefont {G.~J.}\ \bibnamefont {MacDougall}}, \bibinfo {author}
  {\bibfnamefont {M.~J.}\ \bibnamefont {Gilbert}}, \ and\ \bibinfo {author}
  {\bibfnamefont {N.}~\bibnamefont {Mason}},\ }\href
  {https://www.nature.com/articles/s41467-018-05993-w} {\bibfield  {journal}
  {\bibinfo  {journal} {Nature communications}\ }\textbf {\bibinfo {volume}
  {9}},\ \bibinfo {pages} {1} (\bibinfo {year}
  {2018}{\natexlab{b}})}\BibitemShut {NoStop}%
\bibitem [{\citenamefont {Ghatak}\ \emph {et~al.}(2018)\citenamefont {Ghatak},
  \citenamefont {Breunig}, \citenamefont {Yang}, \citenamefont {Wang},
  \citenamefont {Taskin},\ and\ \citenamefont {Ando}}]{ghatak2018anomalous}%
  \BibitemOpen
  \bibfield  {author} {\bibinfo {author} {\bibfnamefont {S.}~\bibnamefont
  {Ghatak}}, \bibinfo {author} {\bibfnamefont {O.}~\bibnamefont {Breunig}},
  \bibinfo {author} {\bibfnamefont {F.}~\bibnamefont {Yang}}, \bibinfo {author}
  {\bibfnamefont {Z.}~\bibnamefont {Wang}}, \bibinfo {author} {\bibfnamefont
  {A.~A.}\ \bibnamefont {Taskin}}, \ and\ \bibinfo {author} {\bibfnamefont
  {Y.}~\bibnamefont {Ando}},\ }\href
  {https://pubs.acs.org/doi/10.1021/acs.nanolett.8b02029} {\bibfield  {journal}
  {\bibinfo  {journal} {Nano letters}\ }\textbf {\bibinfo {volume} {18}},\
  \bibinfo {pages} {5124} (\bibinfo {year} {2018})}\BibitemShut {NoStop}%
\bibitem [{\citenamefont {Yao}\ \emph {et~al.}(2019)\citenamefont {Yao},
  \citenamefont {Xu}, \citenamefont {Wu}, \citenamefont {Aut\`es},
  \citenamefont {Kumar}, \citenamefont {Strocov}, \citenamefont {Plumb},
  \citenamefont {Radovic}, \citenamefont {Yazyev}, \citenamefont {Felser},
  \citenamefont {Mesot},\ and\ \citenamefont {Shi}}]{PhysRevLett.122.176402}%
  \BibitemOpen
  \bibfield  {author} {\bibinfo {author} {\bibfnamefont {M.-Y.}\ \bibnamefont
  {Yao}}, \bibinfo {author} {\bibfnamefont {N.}~\bibnamefont {Xu}}, \bibinfo
  {author} {\bibfnamefont {Q.~S.}\ \bibnamefont {Wu}}, \bibinfo {author}
  {\bibfnamefont {G.}~\bibnamefont {Aut\`es}}, \bibinfo {author} {\bibfnamefont
  {N.}~\bibnamefont {Kumar}}, \bibinfo {author} {\bibfnamefont {V.~N.}\
  \bibnamefont {Strocov}}, \bibinfo {author} {\bibfnamefont {N.~C.}\
  \bibnamefont {Plumb}}, \bibinfo {author} {\bibfnamefont {M.}~\bibnamefont
  {Radovic}}, \bibinfo {author} {\bibfnamefont {O.~V.}\ \bibnamefont {Yazyev}},
  \bibinfo {author} {\bibfnamefont {C.}~\bibnamefont {Felser}}, \bibinfo
  {author} {\bibfnamefont {J.}~\bibnamefont {Mesot}}, \ and\ \bibinfo {author}
  {\bibfnamefont {M.}~\bibnamefont {Shi}},\ }\href {\doibase
  10.1103/PhysRevLett.122.176402} {\bibfield  {journal} {\bibinfo  {journal}
  {Phys. Rev. Lett.}\ }\textbf {\bibinfo {volume} {122}},\ \bibinfo {pages}
  {176402} (\bibinfo {year} {2019})}\BibitemShut {NoStop}%
\bibitem [{\citenamefont {Wang}\ \emph
  {et~al.}(2016{\natexlab{b}})\citenamefont {Wang}, \citenamefont {Gresch},
  \citenamefont {Soluyanov}, \citenamefont {Xie}, \citenamefont {Kushwaha},
  \citenamefont {Dai}, \citenamefont {Troyer}, \citenamefont {Cava},\ and\
  \citenamefont {Bernevig}}]{PhysRevLett.117.056805}%
  \BibitemOpen
  \bibfield  {author} {\bibinfo {author} {\bibfnamefont {Z.}~\bibnamefont
  {Wang}}, \bibinfo {author} {\bibfnamefont {D.}~\bibnamefont {Gresch}},
  \bibinfo {author} {\bibfnamefont {A.~A.}\ \bibnamefont {Soluyanov}}, \bibinfo
  {author} {\bibfnamefont {W.}~\bibnamefont {Xie}}, \bibinfo {author}
  {\bibfnamefont {S.}~\bibnamefont {Kushwaha}}, \bibinfo {author}
  {\bibfnamefont {X.}~\bibnamefont {Dai}}, \bibinfo {author} {\bibfnamefont
  {M.}~\bibnamefont {Troyer}}, \bibinfo {author} {\bibfnamefont {R.~J.}\
  \bibnamefont {Cava}}, \ and\ \bibinfo {author} {\bibfnamefont {B.~A.}\
  \bibnamefont {Bernevig}},\ }\href {\doibase 10.1103/PhysRevLett.117.056805}
  {\bibfield  {journal} {\bibinfo  {journal} {Phys. Rev. Lett.}\ }\textbf
  {\bibinfo {volume} {117}},\ \bibinfo {pages} {056805} (\bibinfo {year}
  {2016}{\natexlab{b}})}\BibitemShut {NoStop}%
\bibitem [{\citenamefont {Borisenko}\ \emph {et~al.}(2019)\citenamefont
  {Borisenko}, \citenamefont {Evtushinsky}, \citenamefont {Gibson},
  \citenamefont {Yaresko}, \citenamefont {Koepernik}, \citenamefont {Kim},
  \citenamefont {Ali}, \citenamefont {van~den Brink}, \citenamefont {Hoesch},
  \citenamefont {Fedorov} \emph {et~al.}}]{borisenko2019time}%
  \BibitemOpen
  \bibfield  {author} {\bibinfo {author} {\bibfnamefont {S.}~\bibnamefont
  {Borisenko}}, \bibinfo {author} {\bibfnamefont {D.}~\bibnamefont
  {Evtushinsky}}, \bibinfo {author} {\bibfnamefont {Q.}~\bibnamefont {Gibson}},
  \bibinfo {author} {\bibfnamefont {A.}~\bibnamefont {Yaresko}}, \bibinfo
  {author} {\bibfnamefont {K.}~\bibnamefont {Koepernik}}, \bibinfo {author}
  {\bibfnamefont {T.}~\bibnamefont {Kim}}, \bibinfo {author} {\bibfnamefont
  {M.}~\bibnamefont {Ali}}, \bibinfo {author} {\bibfnamefont {J.}~\bibnamefont
  {van~den Brink}}, \bibinfo {author} {\bibfnamefont {M.}~\bibnamefont
  {Hoesch}}, \bibinfo {author} {\bibfnamefont {A.}~\bibnamefont {Fedorov}},
  \emph {et~al.},\ }\href {https://www.nature.com/articles/s41467-019-11393-5}
  {\bibfield  {journal} {\bibinfo  {journal} {Nature communications}\ }\textbf
  {\bibinfo {volume} {10}},\ \bibinfo {pages} {1} (\bibinfo {year}
  {2019})}\BibitemShut {NoStop}%
\end{thebibliography}
%

\end{document}